\documentclass[aps,twocolumn,pra,superscriptaddress,showpacs,tightenlines]{revtex4}
\usepackage{amssymb}
\usepackage{amsmath}
\usepackage{graphicx}
\usepackage{epsfig}

\begin{document}

\title{Quantum dynamics of magnetically controlled network for Bloch
electrons }
\author{S. Yang$^{1}$, Z. Song}
\email{songtc@nankai.edu.cn}
\affiliation{Department of Physics, Nankai University, Tianjin 300071, China}
\author{C.P. \surname{Sun}}
\email{suncp@itp.edu.cn} \homepage{http://www.itp.ac.cn/~suncp}
\affiliation{Institute of Theoretical Physics, Chinese Academy of
Sciences, Beijing 100080, China}

\begin{abstract}
We study quantum dynamics of wave packet motion of Bloch electrons
in quantum networks with the tight-binding approach for different
types of nearest-neighbor interactions. For various geometrical
configurations, these networks can function as some optical
devices, such as beam splitters and interferometers. When the
Bloch electrons with the Gaussian wave packets input these
devices, various quantum coherence phenomena can be observed,
e.g., the perfect quantum state transfer without reflection in a
Y-shaped beam, the multi- mode entanglers of electron wave by star
shaped network and Bloch electron interferometer with the lattice
Aharonov-Bohm effects. Behind these conceptual quantum devices are
the physical mechanism that, for hopping parameters with some
specific values, a connected quantum networks can be reduced into
a virtual network, which is a direct sum of some irreducible
subnetworks. Thus, the perfect quantum state transfer in each
subnetwork in this virtual network can be regarded as a coherent
beam splitting process. Analytical and numerical investigations
show the controllability of wave packet motion in these quantum
networks by the magnetic flux through some loops of these
networks, or by adjusting the couplings on nodes. We find the
essential differences in these quantum coherence effects when the
different wave packets enter these quantum networks initially.
With these quantum coherent features, they are expected to be used
as quantum information processors for the fermion system based on
the possible engineered solid state systems, such as the array of
quantum dots that can be implemented experimentally.
\end{abstract}

\maketitle

\section{Introduction}

Quantum information processing (QIP) has been a very active area of research
in the past few years \cite{QIP1,QIP2}. The current challenge for QIP is to
coherently integrate a sufficiently large and complex controllable system
and then requires the ability to transfer quantum information between
spatially separated quantum bits. Since then numerous approaches for this
purpose have been proposed, ranging from linear and nonlinear quantum
optical devices to various interacting quantum systems. Among them, many
studies proposed using the internal dynamics of coupled spins for the
transfer of quantum information \cite{QST1,QST2,QST3,QST4,QST5,LY,ST,SZ,QST6}%
. However, the basic and necessary \textquotedblleft optical
devices\textquotedblright\ (for the electron wave or the spin wave) in a
solid system are scarce due to the technology at hand. Therefore, it is
significant to probe the possibilities to construct the artificial
\textquotedblleft optical devices\textquotedblright\ and then build the
corresponding electronic networks for the matter wave of electron within a
solid state system \cite{quant-network,spin-network,Experiment}. Here, we
notice that, for the boson system, Plenio, Hartley and Eisert \cite{Plenio}\
have studied the quantum network dynamics of a system consisting of a large
number of coupled harmonic oscillators in various geometric configurations
for the similar purpose.

This paper will pay attentions to the fermion systems where the Bloch
electrons move along the quantum lattice network. We consider various
geometrical configurations of tight-binding networks that are analogous to
quantum optical devices, such as beam splitters and interferometers. We then
consider the functions of these tight-binding networks in details when
initially Gaussian wave packets are entering these devices. Analytical and
numerical investigations show that these devices are controllable by the
magnetic flux through the network. Characteristic parameters of devices can
be adjusted by changing the flux or the interactions on nodes. The relevant
quantum phenomena, such as generation of entanglement and the Aharonov-Bohm
(AB) effects in the solid state based devices are also discussed
systematically.

This paper is organized as follows. In section II, we present the
Hamiltonians of the simplest tight-binding lattice systems with and without
magnetic field as building blocks to construct various networks, which can
be formed topologically by the linear and the various connections between
the ends of them. In section III, we theoretically design and analytically
study the basic properties of a star-shaped TBN, then also explore the
further dynamic property of $Y$-shaped network. Surprisingly, for
appropriate joint hopping integrals, the complicated network can be reduced
into an imaginary linear chain with homogeneous NN hopping terms plus a
smaller complicated network. It is known that such TBNs are analogous to
quantum optical devices such as beam-splitters, entangler and
interferometers. In section IV, we investigate the dynamic properties of a
Bloch electron model on a $Q$-shaped lattice, which consists of a terminal
chain and a ring threaded by a magnetic flux. The appropriate flux through
the network can reduce the network to the linear virtual chain, which
indicates that the flux can control the propagation of GWP in the network.
In section V, the interferometer network, a mimic of the AB effect
experiment, is also studied in the similar way. In addition, in the whole
paper, a moving Gauss wave packet (GWP) localized in a linear dispersion
regime is a good example to illustrate the properties of the above Bloch
electron networks. In section VI, we extend the results of the TBNs to the
spin network (SN) for the dynamics of the single magnon. In section VII, we
summarize the results of this paper and suggest the possible applications of
these TBNs.

\section{Basic setup}

In this section, we introduce the systems under consideration, namely the
tight-binding Bloch electron systems and the Hamiltonians of the building
blocks to construct various networks. Without loss of the generality, we
concentrate our attention on the simplest tight-binding systems, in which
only the hopping term or kinetic energy is considered.

A general tight-binding network (TBN) is constructed topologically by the
linear tight-binding chains and the various connections between the ends of
them. An important element in the system is the Aharonov-Bohm flux through
some loops of the TBN. Here, we consider the simplest tight-binding model,
in which only the nearest neighbor (NN) hopping terms are taken into
account. The Hamiltonian of a tight-binding linear chain of $N_{l}$ sites
reads as

\begin{equation}
H_{l}=H_{l}(N_{l})\equiv
-\sum_{j=1}^{N_{l}-1}(t_{j}^{[l]}a_{l,j}^{\dag }a_{l,j+1}+H.c.).
\label{h}
\end{equation}%
Here, the label $l$ denotes the chain with the distribution of the hopping
integrals $\{t_{1}^{[l]},t_{2}^{[l]},...,t_{N_{l}-1}^{[l]}\}$ and $%
a_{l,j}^{\dag }$ is the fermion creation operator at $j$th site of the chain
$l$. The hopping integral could be the complex number due the presence of
the external magnetic field. In this paper, we restrict our study to a
simplest case described as following. When the field is absent, the hopping
integral for a homogeneous chain%
\begin{equation}
t_{j}^{[l]}=te^{i\Phi _{l,j+1}}  \label{ct}
\end{equation}%
in a chain is real and identical (i.e., independent of sites) while in the
presence of a vector potential the hopping integral is modified by a phase
factor. Here, we defined the link phase
\begin{equation}
\Phi _{l,j+1}=\frac{2\pi }{\phi _{0}}\int_{j}^{j+1}\mathbf{A}\cdot d\mathbf{l%
}  \label{phase}
\end{equation}%
with flux quanta $\phi _{0}=hc/e$ as an integral of the vector potential $%
\mathbf{A}$ along the link between the sites $j$ and $j+1$ in the $l$th
chain. The above observation about the phase modification of hopping
integral can be found in many modern literatures \cite%
{gauge-trans,tight-binding} but the proof can be cast back to Peierls {\cite%
{Peierls}}.

Another important portion of the quantum networks is the joints or nodes
between two linear chains, of which the Hamiltonian can be presented in the
form
\begin{equation}
H_{joint}\equiv -(t_{ji}^{[lm]}a_{l,j}^{\dag }a_{m,i}+H.c.),
\label{joint}
\end{equation}%
where $t_{ji}^{[lm]}$\ denotes the hopping integral over the $j$th site of
chain $l$ and the $i$th site of chain $m$. Here, only one joint term
connecting two chains is listed in $H_{joint}$ as an illustration. In a
general TBN, the joint Hamiltonian $H_{joint}$ should contain many
connection terms.

In remaining parts of this paper, we will show that, under certain
conditions, a complex TBNs can be decomposed as a simple sum of several
independent imaginary chains. In order to avoid confusion, we describe one
of the imaginary chains of $N_{l}$ sites by a Hamiltonian%
\begin{equation}
\widetilde{H}_{l}=\widetilde{H}_{l}(N_{l})\equiv -t\sum_{j=1}^{N_{l}-1}(%
\widetilde{a}_{l,j}^{\dag }\widetilde{a}_{l,j+1}+H.c.).
\label{ih}
\end{equation}%
Here, $\widetilde{a}_{l,j}^{\dag }$ denoted by tilde are the fermion
creation operators for $j$th site of the imaginary chain $l$, \ which are
linear combinations of $a_{m,i}$. Namely, there exists a mapping $R$ between
the two sets of fermion operators, $\{a_{l,j}^{\dag }\}\overset{R}{%
\longrightarrow }\{\widetilde{a}_{m,i}^{\dag }\}$ or by a transformation $R$:

\begin{equation}
\widetilde{a}_{m,i}^{\dag }=\sum_{l,j}R_{l,j,m,i}a_{l,j}^{\dag }
\end{equation}%
We will investigate several types of networks based on the notations
introduced above. As an example to demonstrate the application of the
notation, we can express the main conclusion of this paper by using the
above well-defined notations as
\begin{equation}
\sum\limits_{l}H_{l}+H_{joint}\overset{R}{\longrightarrow }\sum\limits_{m}%
\widetilde{H}_{m},
\end{equation}%
i.e., a network can be equivalent to the simple sum of several independent
imaginary chains with the aid of the transformation $R$.

There is a remark to be made here: In usual, the role of the potential $%
\mathbf{A}$ shows as the AB effect of Bloch electron in a close chain (or
called a tight-binding ring). Here, the local magnetic filed strength for
Bloch electron may vanish, but the loop integral of $\mathbf{A}$--the\textbf{%
\ }magnetic flux does not. Due to the AB effect, the magnetic flux $\phi $\
can be used to control the single-particle spectrum of a homogeneous chain, $%
\varepsilon _{k}=-2t\cos (k+2\pi \phi /N)$. When the flux $\phi \sim \phi
_{n}$ $\equiv (n/2$ $+1/4)N$, the lower spectrum becomes a linear dispersion
approximately, i.e., $\varepsilon _{k}\sim k$. For the wave packets as a
superposition of those eigenstates with small $k$, the\ linearization of
Hamiltonian lead to the transfer of the wave packet without spreading \cite%
{YS1}.

\section{The basic blocks of quantum network: star- and Y-shaped beams}

In this section, we use analytical method and numerical simulation to study
the basic blocks of quantum network, the star- and Y-shaped beams, which is
constructed by connecting one ends of several chains to the end (a node) of
a single chain. Beam splitters are the elementary optical devices frequently
used in classical and quantum optics \cite{QOP}, which can even work well in
the level of single photon quanta \cite{S-photon} and are applied to
generate quantum entanglement \cite{Entng}. For matter wave, an early beam
splitter can be referred to the experiments of neutron interference based on
a perfect crystal interferometer with wavefront and amplitude division \cite%
{N-interfer}. Moreover, for cold atoms, a beam splitter have been
experimentally implemented on the atom chip \cite{C-atom}. The theoretical
method has been suggested to realize the beam splitter for the Bose-Einstein
condensate \cite{BEC}.

In the following, we begin with the basic properties of the corresponding
networks for fermions. We will show that, for appropriate joint hopping
integrals, the complicated network can be reduced into an imaginary linear
chain with homogeneous NN hopping terms plus a smaller complicated network.
We also further study the dynamic property of such kind of network by taking
the $Y$-shaped network as an example of star shape beams, in which only
three chains are involved. We investigate various aspects that are analogous
to quantum optical devices, such as beam-splitters, entangler, and
interferometers.

\subsection{Star-shaped beam splitter and its reduction}

\begin{figure}[tbp]
\includegraphics[bb=45 125 560 710, width=7 cm,clip]{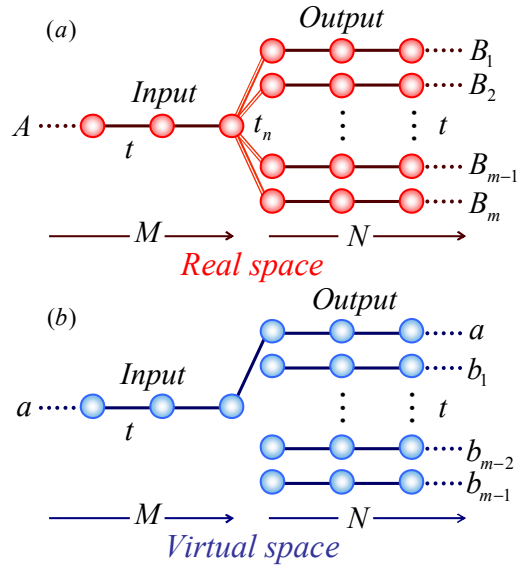}
\caption{\textit{(Color on line) (a) The star-shaped Bloch electron network
with an input chain $A$ and $m$ output chains in the real space. (b) When
the joint hopping constants satisfy $t_{n}=t/\protect\sqrt{m}$, the STBN can
be reduced into one homogeneous tight-binding chain $a$ with length $M+N$
and $m-1$ virtual chains with length $N$ in the virtual space.} }
\label{star}
\end{figure}
We consider a simple TBN, a star shape (we also call the star-shaped
tight-binding network (STBN)) as shown in Fig. \ref{star}(a). The STBN is
constructed by linking the $m$ output chains $B_{p}$ $(p=$ $1,2,$ $...,m)$
to the one end of the input chain\ $A$ by the hopping integrals $%
t_{N_{A}1}^{[AB_{p}]}$. The Hamiltonian of an STBN consists of the linear
chain part (\ref{h}) and the joint part (\ref{joint}) around the last $N_{A}
$th site of the input chain $A$. Obviously, since there is no vector
potential acting on the network, all the hopping integrals are real.

We will show that, under some restriction for the intrachain hopping
constants $t$\ and interchain hopping constants $t_{N_{A}1}^{[AB_{p}]}$, an
STBN can be reduced into an imaginary linear tight-binding chain with
homogeneous hopping constants plus a smaller complicated network. The fact
that the input chain $A$ is a part of this virtual linear chain implies that
the Bloch electron can perfectly propagate in this virtual linear chain
without the reflection by the node. This also indicates that there is a
coherent split of the input electronic wave because the wave function in
this virtual chain actually is just a superposition of wave functions in the
$m$ chains.

To sketch our central idea, we first consider a special STBN, which consists
of $m$ identical \textquotedblleft output\textquotedblright\ chains $%
B_{1},B_{2},\cdots ,$ $B_{m}$ with homogeneous intrachain hopping constants$%
\ t$, interchain hopping constant$\ t_{N_{A}1}^{[AB_{p}]}=t_{n}$ and the
same length $N$, while the length of chain $A$ is $M$. The Hamiltonian of
the network%
\begin{equation}
H=\sum_{p=1}^{m}H_{B_{p}}+H_{A}+H_{joint}
\end{equation}%
is now explicitly written in terms of the chain Hamiltonians $H_{B_{p}}$ and
$H_{A}$ defined by Eq. (\ref{h}). Here, the basic parameters for the network
are $t_{j}^{[B_{p}]}=t_{j}^{[A]}=t$, $N_{B_{p}}=N$, and $N_{A}=M$, where $p=$
$1,2,$ $...,m$. The joint Hamiltonian is
\begin{equation}
H_{joint}=-t_{n}(a_{A,M}^{\dag }\sum_{p=1}^{m}a_{B_{p},1}+H.c.).
\end{equation}%
Note that we only consider the case that all the hopping integrals over the
joints are identical for the convenience of illustration. In the next
section, the different joint hopping integrals will be taken into account
for a simplest case of $m=2$.\

Now we construct the new fermion operators denoted by the tilde notation, $%
\widetilde{a}_{a,j}^{\dag }$\ of virtual tight-binding chain $a$ of length $%
M+N$ as
\begin{align}
\widetilde{a}_{a,j}^{\dag }& =a_{A,j}^{\dag },  \notag \\
\widetilde{a}_{a,M+l}^{\dag }& =\frac{1}{\sqrt{m}}%
\sum_{p=1}^{m}a_{B_{p},l}^{\dag },  \label{EYa}
\end{align}%
where $j\in \lbrack 1,M]$ and $l\in \lbrack 1,N]$. There exist $m-1$
complementary tight-binding chains with the collective operators
\begin{equation}
\widetilde{a}_{b_{q},j}^{\dag }=\frac{1}{\sqrt{m}}\sum_{p=1}^{m}\exp (-i2\pi
pq/m)a_{B_{p},j}^{\dag },  \label{EYb}
\end{equation}%
where $j\in \lbrack 1,N]$, $p=1,2,$ $\cdots ,m,$ and $q=1,2,\cdots ,$ $m-1$.
It can be checked that, all the tilde operators are also the standard
fermion operators, which satisfy the anticommutation relation%
\begin{equation}
\left\{ \widetilde{a}_{\alpha ,i},\widetilde{a}_{\beta ,j}^{\dag }\right\}
=\delta _{\alpha \beta }\delta _{ij},
\end{equation}%
where $\alpha ,\beta \in (a,b_{1},b_{2},\cdots ,b_{m-1})$ denote the labels
of the virtual chains. By inverting Eqs. (\ref{EYa}) and (\ref{EYb}) we have%
\begin{align*}
a_{A,j}^{\dag }& =\widetilde{a}_{a,j}^{\dag },\text{ }(j=1,2,\cdots ,M) \\
a_{B_{p},j}^{\dag }& =\frac{1}{\sqrt{m}}\sum_{q=1}^{m-1}e^{i\frac{2\pi pq}{m}%
}\widetilde{a}_{b_{q},j}^{\dag }+\frac{1}{\sqrt{m}}\widetilde{a}%
_{a,M+j}^{\dag },
\end{align*}%
where $p\in \lbrack 1,m]$, $q\in \lbrack 1,m-1]$, and $j\in \lbrack 1,N]$.\
These establish the mapping $R$ between the two sets of fermion operators, $%
\{a_{l,j}^{\dag }\}\overset{R}{\longrightarrow }\{\widetilde{a}_{m,i}^{\dag
}\}$. Therefore, we have%
\begin{eqnarray}
H &=&-t\sum_{j=1}^{M-1}\widetilde{a}_{a,j}^{\dag }\widetilde{a}%
_{a,j+1}-t\sum_{j=1}^{N-1}(\sum_{q=1}^{m-1}\widetilde{a}_{b_{q},j}^{\dag }%
\widetilde{a}_{b_{q},j+1} \\
&&+\widetilde{a}_{a,M+j}^{\dag }\widetilde{a}_{a,M+j+1})-\sqrt{m}t_{n}%
\widetilde{a}_{a,M}^{\dag }\widetilde{a}_{a,M+1}+H.c..  \notag
\end{eqnarray}

The above Hamiltonian depicts a TBN with different geometry. It is easy to
observe that only when the matching condition of the joint hopping constants
\begin{equation}
t_{N_{A}1}^{[AB_{p}]}=t_{n}=\frac{t}{\sqrt{m}}  \label{march1}
\end{equation}%
is satisfied, we have $H=\widetilde{H}_{a}+\sum_{q=1}^{m-1}\widetilde{H}%
_{b_{q}}:$%
\begin{align}
\widetilde{H}_{a}& =-t\sum_{j=1}^{M+N-1}\widetilde{a}_{a,j}^{\dag }%
\widetilde{a}_{a,j+1}  \notag \\
\widetilde{H}_{b_{q}}& =-t\sum_{j=1}^{N-1}\widetilde{a}_{b_{q},j}^{\dag }%
\widetilde{a}_{b_{q},j+1}+H.c.  \label{reduction}
\end{align}%
where $N_{a}=M+N$ and $N_{b_{q}}=N$. The tilde Hamiltonians are also
illustrated in Fig. \ref{star}(b). Interestingly, all the sub-Hamiltonians $%
\widetilde{H}_{a}$, $\widetilde{H}_{b_{q}}$\ commutate with each other, i.e.,

\begin{equation}
\left[ \widetilde{H}_{\alpha },\widetilde{H}_{\beta }\right] =0,
\end{equation}%
where $\alpha ,\beta \in (a,b_{1},b_{2},\cdots ,b_{m-1})$. This fact means
that the virtual chain described by $\widetilde{H}_{a}$ is just a standard
tight-binding chain of length $N_{a}$ with uniform NN couplings. For an
arbitrary initial state localized within the chain $H_{a}$, it will evolve
driven by the virtual chain of length $M+N$. If the local state moves out of
chain $H_{a}$, an ideal beam splitter can be realized since there is no
reflection occurs at the node.

To demonstrate it, we take an example with the initial state as the Gaussian
wave packet (GWP). The GWP with momentum $\pi /2$ has the form
\begin{equation}
\left\vert \psi (N_{0})\right\rangle =\frac{1}{\sqrt{\Omega }}%
\sum_{j=1}^{M}e^{-\frac{^{\alpha ^{2}}}{2}(j-N_{0})^{2}}e^{i\frac{\pi }{2}%
j}\left\vert j\right\rangle .  \label{GWP}
\end{equation}%
Here, $\Omega =\sum_{j=1}^{N_{A}}\exp [-\alpha ^{2}(j-N_{0})^{2}]$ is the
normalization factor and $N_{0}\in \lbrack 1,N_{A}]$ is the initial central
position of the GWP at the input chain $A$, while the factor $\alpha $ is
large enough to guarantee the locality of the state in the chain $A$.
Accordingly, the basis $\left\vert j\right\rangle $ is defined as $%
\left\vert j\right\rangle =\widetilde{a}_{a,j}^{\dag }\left\vert
0\right\rangle $ for $j\in \lbrack 1,M+N]$. In the previous work \cite{YS1},
it has been shown that such GWP can approximately propagate along a
homogenous chain without spreading. Actually, at a certain time $\tau $,
such GWP evolves into%
\begin{eqnarray}
\left\vert \Psi (\tau )\right\rangle &=&e^{-i\widetilde{H}_{a}\tau
}\left\vert \psi (N_{0})\right\rangle \simeq \left\vert \psi (N_{0}+2t\tau
)\right\rangle  \notag \\
&=&\frac{1}{\sqrt{\Omega }}\sum_{j=M+1}^{M+N}e^{-\frac{^{\alpha ^{2}}}{2}%
(j-N_{0}-2t\tau )^{2}}e^{i\frac{\pi }{2}j}\left\vert j\right\rangle
\end{eqnarray}%
in the virtual space. From the mapping of the operators (\ref{EYa}), we have
the final state as%
\begin{equation}
\left\vert \Psi (\tau )\right\rangle =\frac{1}{\sqrt{m}}\sum_{p=1}^{m}\left%
\vert \phi _{p}(N_{\tau })\right\rangle .
\end{equation}%
Here, the state

\begin{equation}
\left\vert \phi _{p}(N_{\tau })\right\rangle =\frac{1}{\sqrt{\Omega }}%
\sum_{j=1}^{N}e^{-\frac{^{\alpha ^{2}}}{2}(j-N_{\tau })^{2}}e^{i\frac{\pi }{2%
}j}a_{B_{p},j}^{\dag }\left\vert 0\right\rangle ,
\end{equation}%
is the clone of the initial GWP with the center at $N_{\tau }=N_{0}+2t\tau
-M $ in the chain $B_{p}$. Then we conclude that the beam splitter split the
single-particle GWP into $m$ cloned GWPs without any reflection.
Furthermore, it is obvious that the function of the splitter originates from
the reduction (\ref{reduction}) of the Hamiltonian, which is available for
every invariant subspaces of fixed particle number. Therefore, such splitter
can be applied to the many-particle system.

The above discussion is limited to the simplest case of identical joint
hopping integrals. We would like to say that the marching condition (\ref%
{march1}) is not unique for constructing independent virtual chain. We will
demonstrate this for the case $m=2$ in the next section.

\subsection{Y-shaped beam splitter}

\begin{figure}[tbp]
\includegraphics[bb=50 210 550 660, width=7 cm,clip]{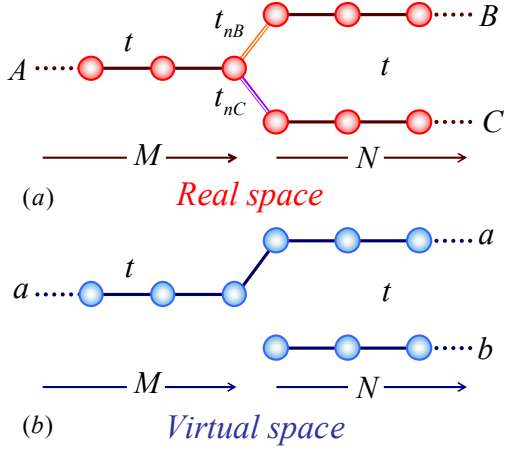}
\caption{\textit{(Color on line) (a) $Y$-shaped TBN or called $Y$-beam, a
special STBN with different joint hopping integrals $t_{nB}$ and $t_{nC}$.
(b) Reduction of $Y$-shaped TBN under the matching condition. It shows that
if $t_{nB}^{2}+t_{nC}^{2}=t^{2}$, the $Y$-shaped TBN in real space is mapped
into virtual space as two decoupled virtual chain $a$ and $b$ with length $%
M+N$ and $N$ respectively.}}
\label{Y}
\end{figure}
The reduction for star-shaped network was demonstrated under the
restriction\ (\ref{march1}). This kind of reduction can be also performed
with different joint hopping integrals. If the splitter is applied to the
local GWP, the node interactions of the two output legs are not necessary to
be identical. In the following, we will investigate this issue by
considering the simplest configuration with $m=2$, which is called $Y$-beam.
The asymmetric $Y$-beam consists of three legs $A$, $B$ and $C$ with the
intrachain hopping integrals $t$ for $F=A,B,C$ and the joint ones $t_{nF}$
for $F=B,C$ (see Fig. \ref{Y}(a)). The total Hamiltonian reads
\begin{equation}
H=\sum_{F=A,B,C}H_{F}-\sum_{F=B,C}(t_{nF}a_{A,M}^{\dag
}a_{F,1}+H.c.)
\end{equation}%
where $t_{j}^{[A]}=t_{j}^{[B]}$ $=t_{j}^{[C]}=t$, $N_{A}=M,$ and $%
N_{B}=N_{C}=N$.

In order to decouple this $Y$-beam\emph{\ }as two virtual linear
tight-binding chains, we need to optimize the asymmetric couplings so that
the perfect transmission can occur in the decoupled linear tight-binding
chains. For this purpose, we introduce the tilde operators of fermion%
\begin{align}
\widetilde{a}_{a,j}^{\dag }& =a_{A,j}^{\dag },  \notag \\
\widetilde{a}_{a,M+l}^{\dag }& =\cos \theta a_{B,l}^{\dag }+\sin \theta
a_{C,l}^{\dag },  \notag \\
\widetilde{a}_{b,m}^{\dag }& =\sin \theta a_{B,m}^{\dag }-\cos \theta
a_{C,m}^{\dag },  \label{tilde_Y}
\end{align}%
where $j\in \lbrack 1,M]$, $l\in \lbrack 1,N]$,\ and $m\in \lbrack 1,N]$.
Here, the mixing angle $\theta $ is to be determined as follows by the
optimization for quantum state transmission. In comparison with the optical
beam splitter, the above equations (\ref{tilde_Y}) can be regarded as a
fundamental issue for the electronic wave beam splitter.
\begin{figure}[tbp]
\includegraphics[bb=30 240 500 650, width=7 cm,clip]{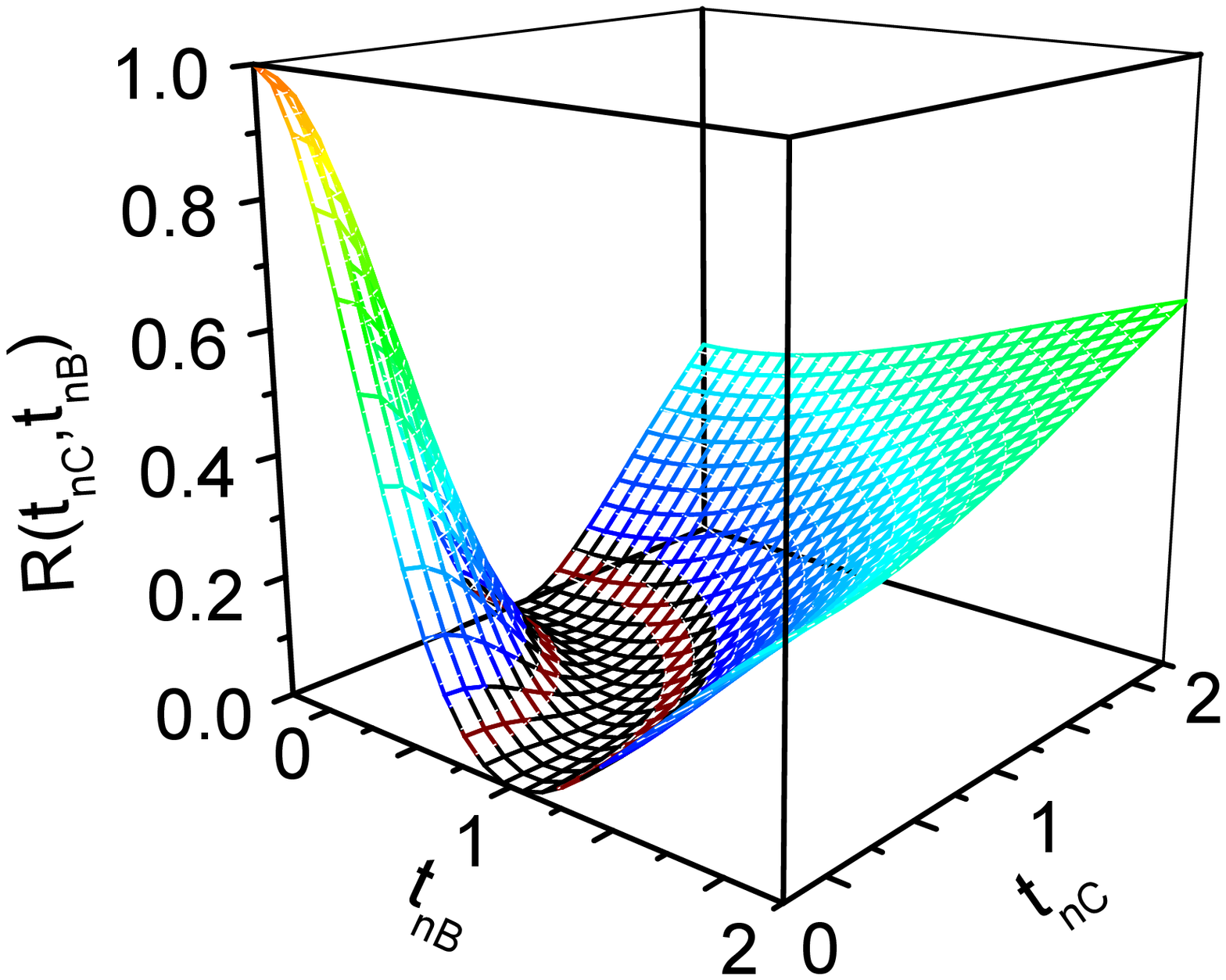} %
\includegraphics[bb=20 315 490 785, width=6.5 cm,clip]{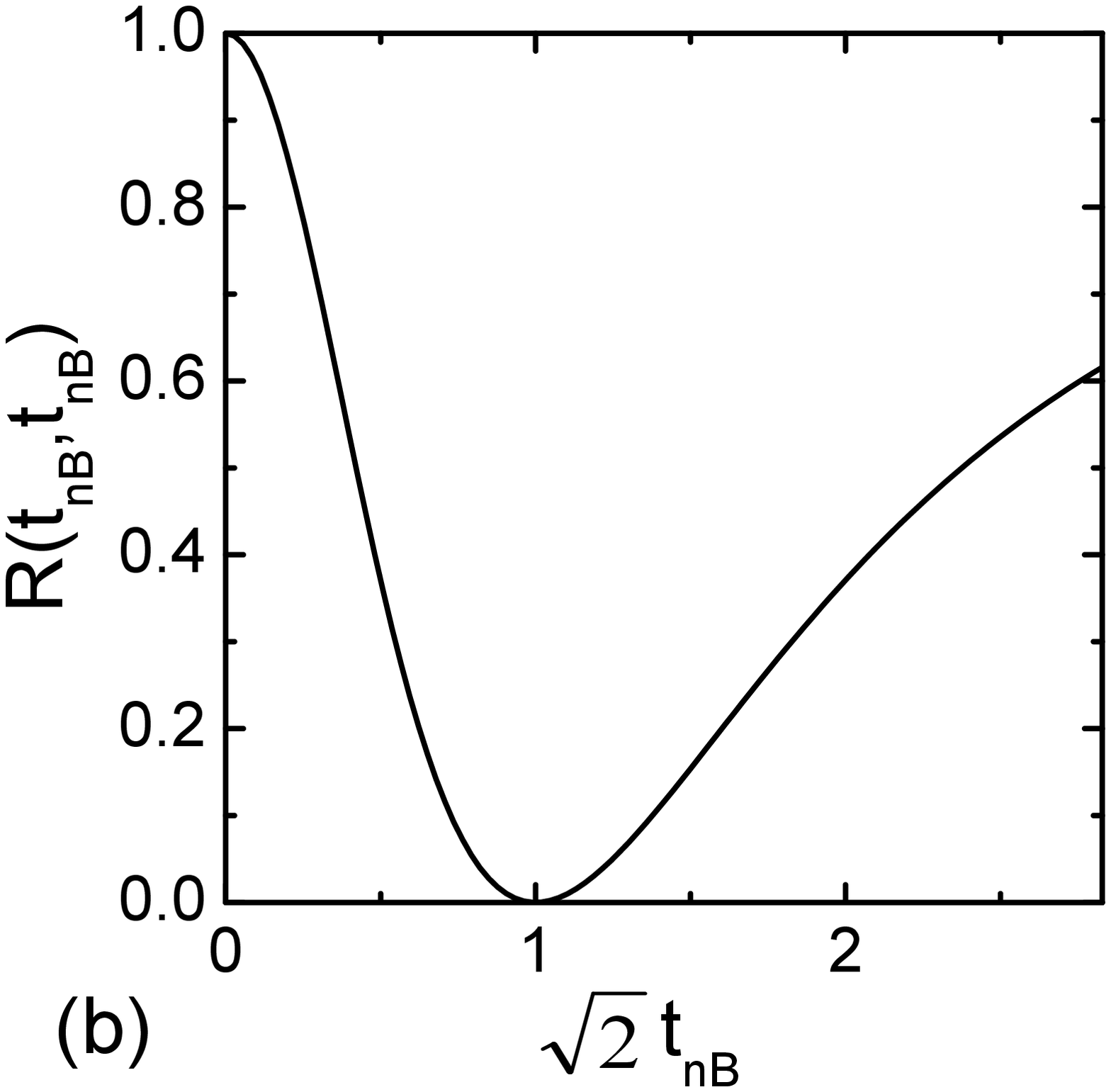}
\caption{\textit{(Color on line) (a) The contour map of the reflection
factor $R(t_{nC},t_{nB})$ as a function of $t_{nC},t_{nB}$ for the GWP with $%
\protect\alpha =0.3$ and momentum $\protect\pi /2$ in a finite system with $%
N_{A}=$$N_{B}=$$N_{C}=50$. It shows that around the matching condition, i.e,
the circle $t_{nC}^{2}+t_{nB}^{2}=t^{2}$, the reflection factor approaches
zero. (b) The profile of $R(t_{nC},t_{nB})$ along $t_{nC}=t_{nB}$. }}
\label{R}
\end{figure}

Together with the original creation operator $\widetilde{a}_{a,j}^{\dag
}=a_{A,j}^{\dag }$ for the input leg, the set $\{\widetilde{a}_{a,M+j}^{\dag
}\mid j\in \lbrack 1,N]\}$ defines a new linear chain $a$ with the effective
hopping integrals
\begin{eqnarray}
t_{aj} &=&t_{a,M+l}=t,  \notag \\
t_{aM} &=&t_{nB}\cos \theta +t_{nC}\sin \theta ,
\end{eqnarray}%
where $j\in \lbrack 1,M-1]$\ and $l\in \lbrack 1,N-1]$.\ Another virtual
linear chain $b$ is defined by $\widetilde{a}_{b,j}^{\dag }$ with
homogeneous hopping integral $t_{bj}=t$,$\ j\in \lbrack 1,N-1]$.

In general, these two linear chains are dependent, since there is a
connection interaction around the node
\begin{equation}
H_{joint}=-t_{AB}(\widetilde{a}_{a,M}^{\dag
}\widetilde{a}_{b,1}+H.c.)
\end{equation}%
where
\begin{equation}
t_{AB}=t_{nB}\sin \theta -t_{nC}\cos \theta .
\end{equation}%
Fortunately, the two virtual chains decouple with each other when the mixing
angle $\theta $ and the intrachain connections are optimized by setting $%
\tan \theta =t_{nC}/t_{nB}$. And if we take $t_{AB}=t$, the virtual chain $a$%
\ becomes a completely homogeneous chain of length $M+N$ as illustrated in
Fig. \ref{Y}(b).\ Thus, these lead to the matching\ condition%
\begin{equation}
\sqrt{t_{nC}^{2}+t_{nB}^{2}}=t  \label{matching}
\end{equation}%
for $Y$-beam network. It can be employed to transfer the quantum state
without reflection on the node in the transformed picture. Transforming back
to the original picture, we can see that such network behaves as a perfect
beam splitter.

In the point of view of linear optics, such beam splitting process can
generate the\ mode entanglement between the separated waves in chains $B$
and $C$, and the measure of such mode entanglement is determined by the
values of $t_{nB}$ and $t_{nC}$. We will show that the strength of $t_{nB}$
and $t_{nC}$ can be used to control the amplitudes of the evolving Bloch
electron wave packets on legs $B$ and $C$.

Now we apply the beam splitter to a special Bloch electron wave packet, a
GWP with momentum $\pi /2$, which has the form (\ref{GWP}) at $\tau =0$. It
is known from the previous work \cite{YS1} that such GWP can approximately
propagate along a homogenous chain without spreading. Then at a certain time
$\tau $, such GWP evolves into
\begin{equation}
\left\vert \Psi (\tau )\right\rangle =\cos \theta \left\vert \psi
_{B}(N_{\tau })\right\rangle +\sin \theta \left\vert \psi _{C}(N_{\tau
})\right\rangle  \label{split}
\end{equation}%
where
\begin{equation}
N_{\tau }=N_{0}+2t\tau -M,
\end{equation}%
i.e., the beam splitter can split the GWP into two cloned GWPs completely.
The possibility of GWPs in the arms $B$ and $C$ is determined by the mixed
angle $\theta $.

In order to verify the above analysis, a numerical simulation is performed
for a GWP with $\alpha =0.3$ in a finite system with $N_{A}=N_{B}$ $=$$%
N_{C}=50$. Let $\left\vert \Phi (0)\right\rangle $ be a normalized initial
state. Then the reflection factor at time $\tau $ can be defined as%
\begin{equation}
R(t_{nC},t_{nB},\tau )=\sum_{j_{A}=1}^{M-1}\left\vert \left\langle
j_{A}\right\vert e^{-iH\tau }\left\vert \Phi (0)\right\rangle \right\vert
^{2}
\end{equation}%
to depict the reflection at the node. At an appropriate instant $\tau _{0}$,
\begin{equation}
R(t_{nC},t_{nB})=R(t_{nC},t_{nB},\tau _{0})
\end{equation}%
as a function of $t_{nC}$ and $t_{nB}$ is plotted in Fig. \ref{R}. It shows
that around the matching condition (\ref{matching}), the reflection factor
vanishes, which is just in agreement with our analytical result.

The conclusion for such GWP comes from the reduction of the $Y$-shaped
network, which the equal length of two output arms is crucial. However, for
the local wave packet (\ref{GWP}), the local environment around the wave
packet only result in its behavior at the next instant. This can be seen
from the speed of the GWP (\ref{GWP}). According to the study in Ref. \cite%
{YS1}, the speed of the GWP is independent of the size and the boundary
condition (ring or open chain). Therefore, for a splitter to GWP, the
equality of two output arms is not necessary. This argument will be
demonstrated in the following content about quantum interferometer.

\begin{figure}[tbp]
\includegraphics[bb=30 240 500 650, width=7 cm,clip]{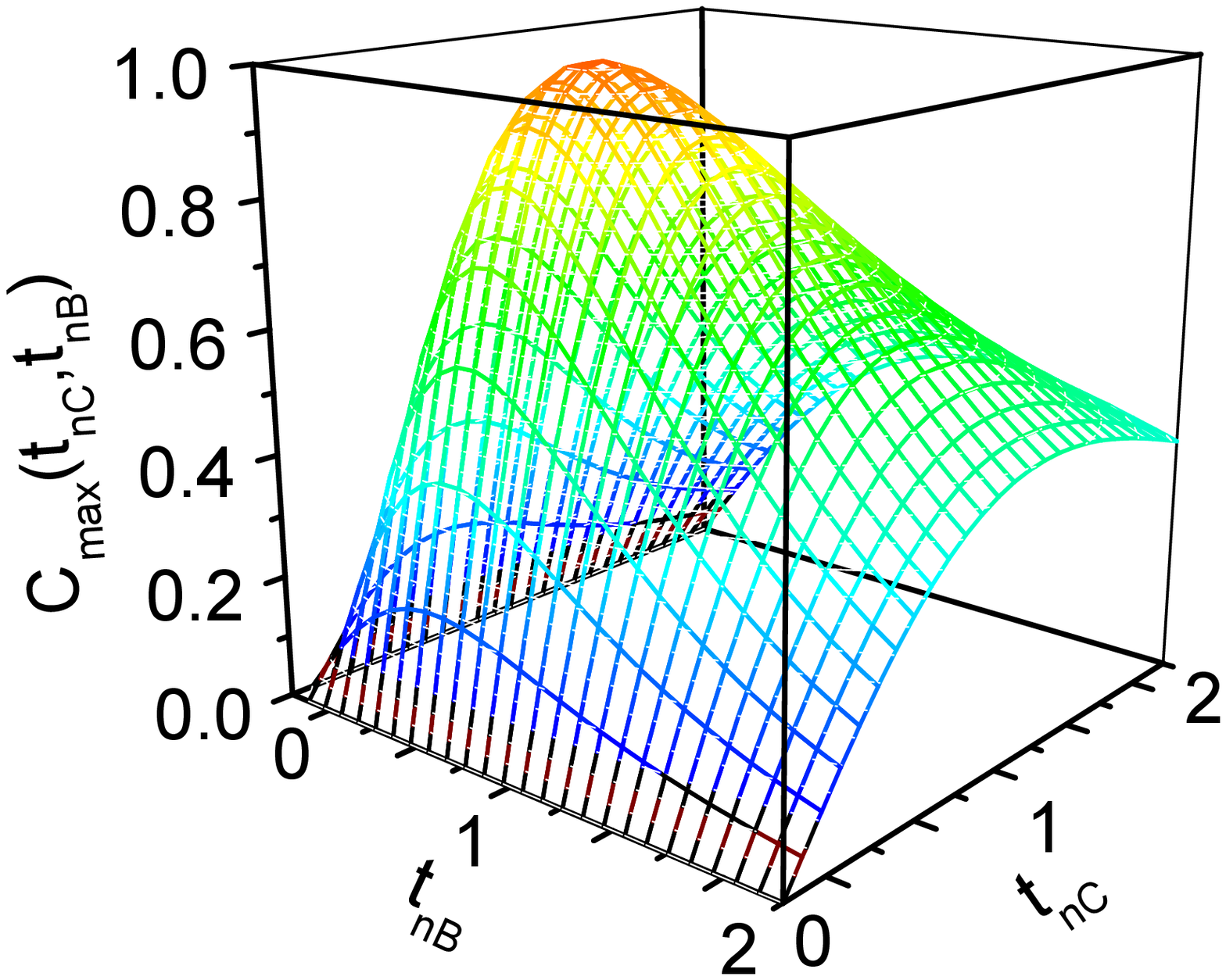} %
\includegraphics[bb=15 315 490 785, width=6.5 cm,clip]{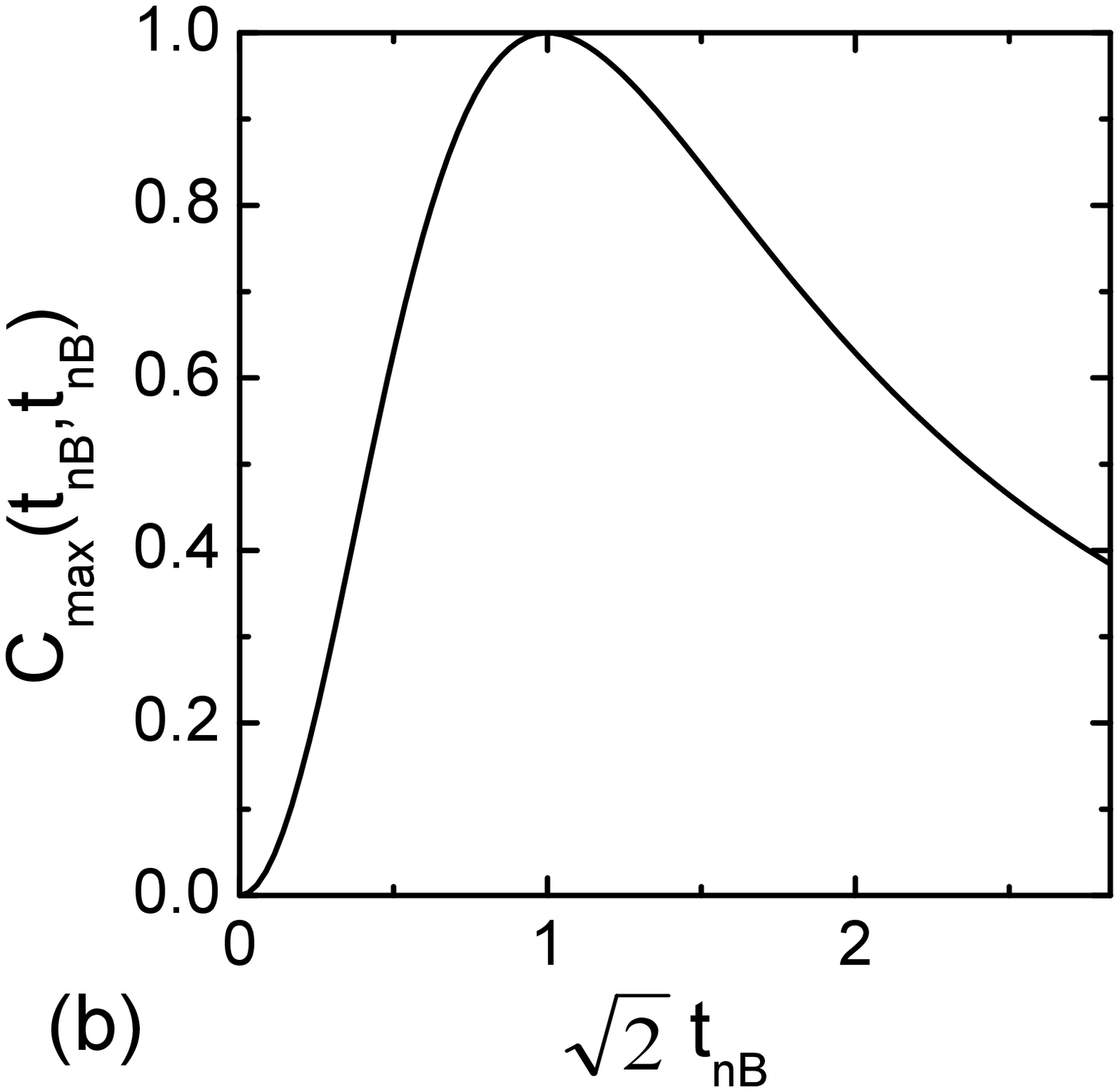}
\caption{\textit{(Color on line) (a) The contour map of maximal concurrence
of two GWPs at two legs $A$ and $B,$ $C_{\max }(t_{nC},t_{nB})$ for the same
setup as that in Fig. \protect\ref{R}. It is found that two GWPs yield the
maximal entanglement at the point $t_{nC}=$$t_{nB}=$$t_{A}/\protect\sqrt{2}$%
. (b) The profile of $C_{\max }(t_{nC},t_{nB})$ along $t_{nC}=t_{nB}$.}}
\label{C}
\end{figure}

\subsection{Entangler of Bloch electron}

Now we consider how the STBN can behave as an entangler to produce
entanglement with the $Y$-beam\textbf{\ }as an illustration. Let the input
state $|\phi (0)\rangle $, represents single-particle state located in the
arm $A$. It can propagate into the arms $B$ and $C$ through the node with
some reflection. On the other hand, the electronic wave can be regarded as
being transferred along the virtual legs $a$ and $b$. Once we manipulate the
joint hopping integrals to satisfy the matching condition, the electronic
wave can only enter the virtual chain $a$ rather than $b$ without any
reflection. Then the final state is in the subspace of the virtual chain $a$%
. Similar to optical splitter, such $Y$-beam splitting can also be regarded
as an entangler of fermion. For instance, consider a state $|D(j)\rangle =%
\widetilde{a}_{a,j}^{\dag }|0\rangle $ for $j\in \lbrack M,M+N]$, which is a
local state in the view of point of the virtual chain $a$. However, in the
real space, this state is nonlocal and possesses mode entanglement, while
state $|D(j)\rangle $ for $j\in $ $[1,M]$ is still a non-entangled state.
Obviously, the $Y$-beam acts as an entangler similar to that in quantum
optical systems.

To quantitatively characterize mode entanglement generated by the splitter
on the joint hopping integrals $t_{nC}$, $t_{nB}$, we consider the GWP (\ref%
{GWP}) as an initial state. Through the splitter, two separated GWPs are
obtained. The total concurrence with respect to the two waves located at the
arms $B$ and $C$ can be calculated as
\begin{equation}
C(\tau )=\sum_{j=1}^{N}\left\vert \left\langle \Psi (\tau )\right\vert
(a_{B,j}^{\dag }a_{C,j}+a_{B,j}a_{C,j}^{\dag })\left\vert \Psi (\tau
)\right\rangle \right\vert
\end{equation}%
according to refs.\cite{wang,qian}. When the interchain connections are
optimized by setting $t_{nB}=t\cos \theta $, $t_{nc}=t\sin \theta $, the
above mode concurrence can be calculated as%
\begin{equation}
C(\tau )=\sin (2\theta )
\end{equation}%
from the Eq. (\ref{split}).

It is obvious that if $\cos \theta =$ $\sin \theta =1/\sqrt{2}$, i.e.,
\begin{equation}
t_{nB}=t_{nC}=\frac{t}{\sqrt{2}},  \label{MaxC}
\end{equation}%
$C(\tau )$ reaches its maximum value $1$. Numerical simulation for $%
t_{nB},t_{nC}\in \lbrack 0,2t]$ is performed for a GWP with $\alpha =0.3$
and momentum $\pi /2$ in a finite system with $N_{A}=50,$ $N_{B}=50,$ and $%
N_{C}=50$. The concurrence is also the function of time $\tau $\ due to the
dynamics of the system. We choose maximal concurrence
\begin{equation}
C_{\max }(t_{nC},t_{nB})=\max \{C(\tau )\}
\end{equation}%
as a function of $t_{nC}$ and $t_{nB}$ to depict the property of the
splitter. Numerical result is plotted in Fig. \ref{C}. It shows that two
split wave packets yield the maximal entanglement just at the matching point
(\ref{MaxC}).

\subsection{Quantum interferometer for Bloch electron}

\begin{figure}[tbp]
\includegraphics[bb=45 330 520 625, width=7 cm,clip]{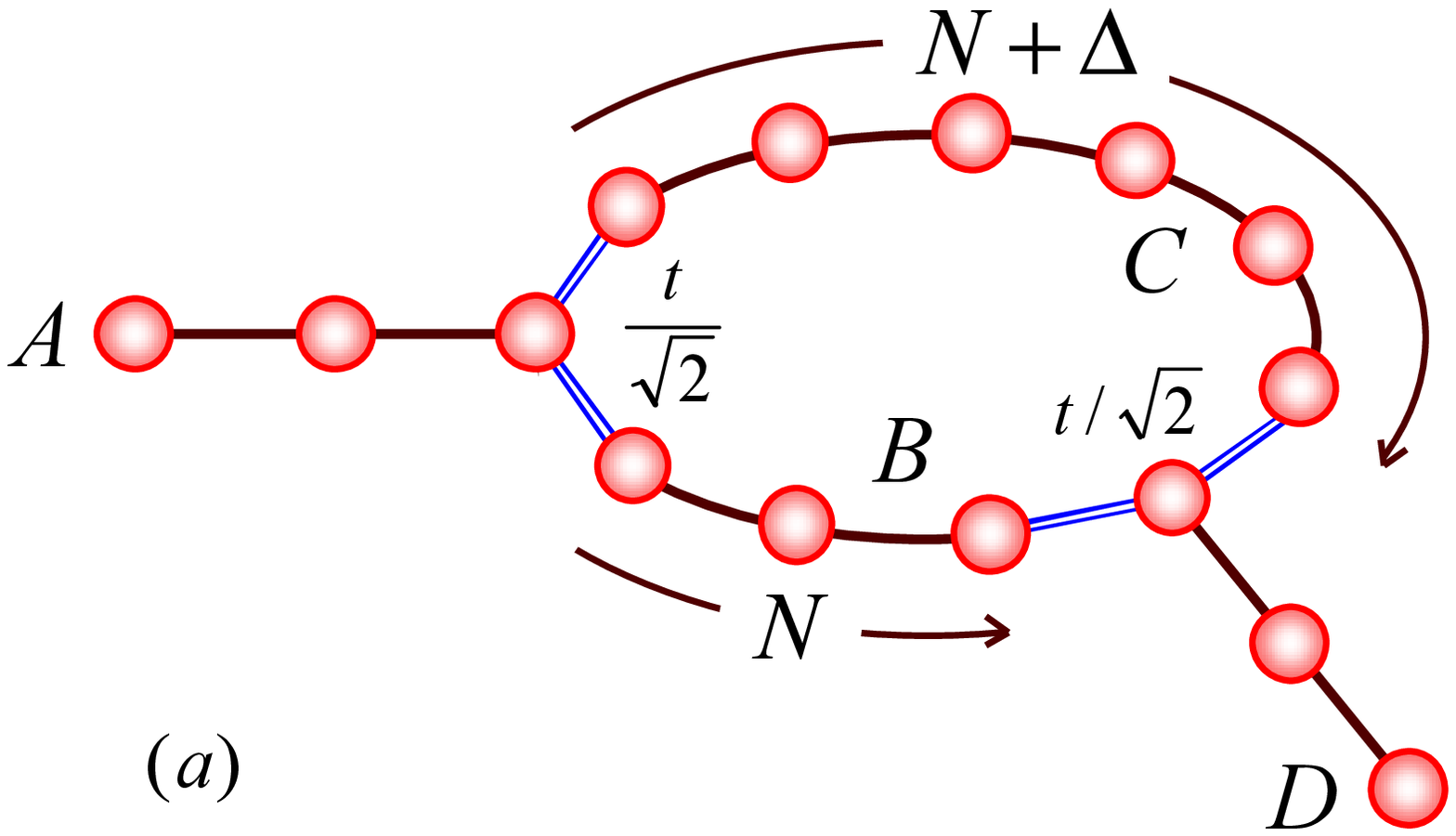} %
\includegraphics[bb=40 445 530 690, width=7 cm,clip]{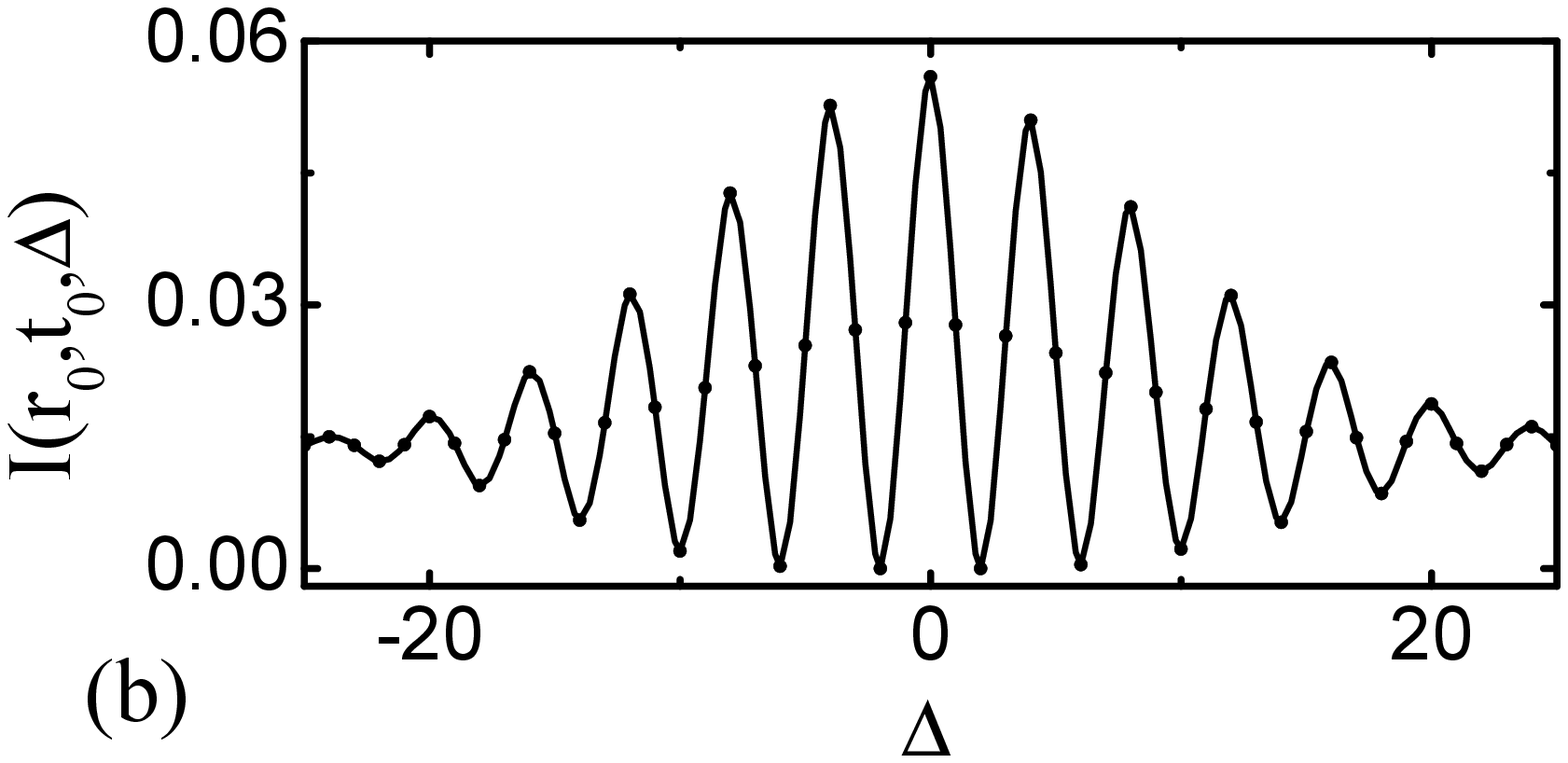}
\caption{\textit{(Color on line) (a) The interferometric network with an
input chain $A$ and output chain $D$, which consists of two $Y$-beams. $%
\Delta $ is the \textquotedblleft optical path difference\textquotedblright\
which determines the interference pattern of output spin wave. (b) The
interference pattern of output wave in the leg $D$ ($r_{0}=50$, $%
t_{0}=100/J_{A}$) for the GWP with $\protect\alpha =0.3$ in the
interferometric network with $N_{A}=N_{B}$ $=N_{D}=50$, $N_{C}=N_{B}+\Delta $%
. }}
\label{inter}
\end{figure}

Now we consider in details a more complicated TBN than the $Y$-beam, the
quantum interferometer for Bloch electron wave. This setup consists of two $%
Y $-beams, which is illustrated schematically in Fig. \ref{inter}(a). It is
similar to the optical interferometer, where state $\left\vert
a\right\rangle $\ of a single photon is split into two parts $\left\vert
b\right\rangle $ and $\left\vert c\right\rangle $ by the splitter and then a
new state $\left\vert d\right\rangle =U_{B}\left\vert b\right\rangle
+U_{C}\left\vert c\right\rangle $ can be achieved by the unitary
transformations $U_{B}$\ and $U_{C}$. In the tight-binding Bloch electron
interferometer, the analogue of the import state $\left\vert a\right\rangle $%
\ is the moving GWP (\ref{GWP}).

Firstly, we consider the simplest case with the path difference (defined in
Fig. \ref{inter}(a)) $\Delta =0$. It is shown that such network is
equivalent to two independent virtual chains with lengths $N_{A}+N_{B}+N_{D}$
and $N_{B}$ respectively when the coupling matching condition is satisfied.
Then the initial GWP will be transmitted into the arm $D$ without any
reflection. This fact can be understood according to the interference of two
split GWPs. Actually, from the above analysis about the GWP propagating in
the $Y$-beam, we note that the conclusion can be extended to the $Y$-beam
consisting of two different length arms $N_{B}\neq N_{C}$. It is due to the
locality of the GWP and the fact that the speed of the GWP only depends on
the hopping integral. Then the arrival time of the two split GWPs at arm $D$
depends on the lengths $N_{B}\ $and $N_{C}$.\ It means that the nonzero $%
\Delta $ should affect the shape of the pattern of output wave.

To verify the analysis above, we investigate this problem again numerically.
According to quantum mechanics, the interference pattern at site $r_{0}$ and
time $\tau _{0}$ in arm $D$ can be presented as
\begin{equation}
I(r_{0},\tau _{0},\Delta )=\left\vert \left\langle r_{0}\right\vert \exp
(-iH\tau _{0})\left\vert \Phi (0)\right\rangle \right\vert ^{2}.  \label{I}
\end{equation}%
Numerical simulation of $I(r_{0},\tau _{0},\Delta )$ for the input GWP in
the interferometric network with $N_{A}=N_{B}$ $=N_{D}=50$, $%
N_{C}=N_{B}+\Delta $\ is performed. For $r_{0}=50$, $\tau _{0}=100/t$, a
perfect interference phenomenon by $I(r_{0},\tau _{0},\Delta )$ is observed
for the range $\Delta \in \lbrack -25,25]$ in Fig. \ref{inter}(b). This
observation shows that the quantum interferometer can be realized by the TBN.

\section{$Q$-shaped TBN controlled by flux}

From the above discussion, it can be found that the essence of the reduction
for the TBN lies on the interference of the matter wave. On the other hand,
the presence of vector potential can induce a phase factor to the wave
function and then the magnetic flux can control the coherent reduction to
some extent. In this section, we investigate how to control the motion of
the Bloch electron along this TBN by an external magnetic field. We will
show that the appropriate flux through the network can reduce the network to
the linear virtual chain, which indicate that the flux can control the
propagation of GWP in the network.

\subsection{Model and Hamiltonian}

\begin{figure}[tbp]
\includegraphics[bb=60 100 535 760, width=7 cm,clip]{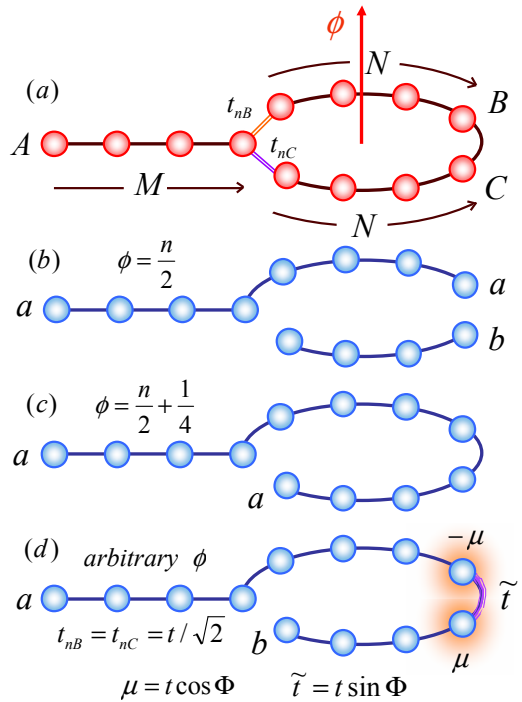}
\caption{\textit{(Color on line) (a) The $Q$-shaped Bloch electron network
with an input chain $A$ and a ring ${B,C}$ threaded by a magnetic flux. (b)
When $\protect\phi =n/2$ and $t_{nB}=t_{nC}=t/\protect\sqrt{2}$, the $Q$%
-shaped Bloch electron network can be decoupled into two virtual homogeneous
linear chains $a$ and $b$ with length $M+N$ and $N$ respectively. (c) When $%
\protect\phi =n/2+1/4$ and $t_{nB}^{2}+t_{nC}^{2}=t^{2}$, the $Q$-shaped
Bloch electron network can be decoupled into a long virtual homogeneous
linear chain $a$ with length $M+2N$. (d) If $t_{nB}=t_{nC}=t/\protect\sqrt{2}
$, as to an arbitrary $\protect\phi $, the virtual homogeneous linear chains
$a$ and $b$ are connected by the hopping integral $\widetilde{t}$. There
also exist chemical potentials $-\protect\mu $ and $\protect\mu $ at the
ends of the virtual chains $a$ and $b$ respectively.}}
\label{Q}
\end{figure}

Consider a quantum network constructed by connecting the two free ends of
the two identical chains in $Y$-beam. Such a network is called $Q$-shaped
TBN, or QTBN labelled by $\{A,B,C\}$. As illustrated schematically in Fig. %
\ref{Q}(a), QTBN is placed in an external magnetic field. The ring of the
model is threaded by a magnetic flux $\phi $ in the unit of flux quanta.
Here, we only consider effect of the vector potential $\mathbf{A}$\ without
the Zeeman term for simply. The Hamiltonian of our $Q$-shaped lattice model
reads
\begin{equation}
H=H_{A}+H_{B}+H_{C}+H_{joint}  \label{H-Q}
\end{equation}%
where
\begin{eqnarray}
H_{joint} &=&-(t_{nB}a_{A,M}^{\dag }a_{B,1}e^{i\Phi
_{B,1}}+t_{nC}a_{A,M}^{\dag }a_{C,1}e^{-i\Phi _{C,1}})  \notag \\
&&-ta_{B,N}^{\dag }a_{C,N}e^{i\Phi _{BC}}+H.c.
\end{eqnarray}%
and the parameters $N_{A}=M$, $N_{B}=N_{C}=N$, $t_{j}^{[A]}=t$, $%
t_{j}^{[B]}=t\exp (i\Phi _{B,j+1})$, $t_{j}^{[C]}=t\exp (-i\Phi _{C,j+1})$.
Here, $\Phi _{B,l}$, $\Phi _{C,l}$, $l\in \lbrack 1,,N]$ denote the phase
differences between the neighboring sites $l$ and $l+1$ in the chains $B$
and $C$, while $\Phi _{BC}$ is respect to the connection between the two
chain. The values of the phase difference is defined as (\ref{phase}) and $%
\Phi _{B,l}$, $\Phi _{C,l}$, and $\Phi _{BC}$\ are not required to be
identical in order to avoid losing generality.

In the following discussion, what we concern is only the sum of the phase
difference along the loop%
\begin{equation}
\Phi =\sum_{l=1}^{N}(\Phi _{B,l}+\Phi _{C,l})+\Phi _{BC}
\end{equation}%
corresponding to the flux $\phi =\Phi /2\pi $\ through the loop. We will
show that the flux $\phi $\ can control the dynamics of the $Q$-shaped
lattice system.

\subsection{Reduction of the $Q$-shaped lattice model}

The QTBN with the flux $\phi $, and the joint hopping strengths $t_{nB}$, $%
t_{nC}$, exhibits a rich variety of dynamic behaviors. Fortunately, we find
that there exist analytical results in some range of parameters. Together
with the numerical simulation, these analytical results are helpful to get a
comprehensive understanding of the mechanism. We start with the cases of
fixed $\Phi $ but various $t_{nB}$, $t_{nC}$ and then investigate the cases,
vice versa.%

\subsubsection{Case: $\protect\phi =\frac{n}{2}$}

Our aim is try to decouple this $Q$-shaped model as two virtual linear
chains.\ We first introduce two anticommutative sets of fermion operator $\{%
\widetilde{a}_{a,M+j}^{\dag }\}$, $\{\widetilde{a}_{b,j}^{\dag }\}$\ defined
by%
\begin{eqnarray}
\widetilde{a}_{a,M+j}^{\dag } &=&\cos \theta e^{-i\varphi
_{B}^{j}}a_{B,j}^{\dag }+\sin \theta e^{i\varphi _{C}^{j}}a_{C,j}^{\dag }
\notag \\
\widetilde{a}_{b,j}^{\dag } &=&\sin \theta e^{-i\varphi
_{B}^{j}}a_{B,j}^{\dag }-\cos \theta e^{i\varphi _{C}^{j}}a_{C,j}^{\dag }
\label{tildeQ}
\end{eqnarray}%
for $j\in \lbrack 1,N]$, where $\varphi _{\alpha }^{j}=\sum_{l=1}^{j}\Phi
_{\alpha ,l}$, $(\alpha =B,C)$.

We can check that they still satisfy the anticommutation relations $\{%
\widetilde{a}_{\alpha ,i},\widetilde{a}_{\beta ,j}^{\dag }\}=\delta _{\alpha
\beta }\delta _{ij}$, where $\alpha ,\beta \in (a,b)$. The inverse
transformations of Eq. (\ref{tildeQ}), together with the original fermion
operator $\widetilde{a}_{a,j}^{\dag }=a_{A,j}^{\dag },$ $j\in \lbrack 1,M]$,
define a new linear chain $a$, while another virtual linear chain $b$ is
only constructed by $\widetilde{a}_{b,j}^{\dag }$, $j\in \lbrack 1,N]$.
Therefore, the parameters are taken as

\begin{equation}
\phi =\frac{n}{2};t_{nB}=t_{nC}=\frac{t}{\sqrt{2}},
\end{equation}%
the Hamiltonian can be reduced as%
\begin{eqnarray}
H &=&\widetilde{H}_{a}+\widetilde{H}_{b}+\widetilde{H}_{\mu }  \notag \\
\widetilde{H}_{\mu } &=&t(-1)^{n}\widetilde{n}_{a,M+N}-t(-1)^{n}\widetilde{n}%
_{b,N}
\end{eqnarray}%
with $N_{a}=M+N$ and $N_{b}=N$.\ Here, $\widetilde{n}_{a,M+N}=\widetilde{a}%
_{a,M+N}^{\dag }\widetilde{a}_{a,M+N}$ and $\widetilde{n}_{b,N}=\widetilde{a}%
_{b,N}^{\dag }\widetilde{a}_{b,N}$ are the particle number operators. The $%
\widetilde{H}_{\mu }$\ represents the chemical potential at the ends of
chains $a$ and $b$. For large $N$ system, the effect of the end potentials
can be ignored. Thus the $Q$-type lattice can be reduced into two
independent virtual linear chains $a$ and $b$ with homogeneous NN hopping
integrals, and length $M+N$ and $N$ respectively as illustrated in Fig. \ref%
{Q}(b).\

\subsubsection{Case: $\protect\phi =\frac{n}{2}+\frac{1}{4}$}

In this case, we will show that the model can be reduced to a virtual chain
with $M+2N$ sites if the joint hopping integrals satisfy $\sqrt{%
t_{nC}^{2}+t_{nB}^{2}}=t$. We introduce the fermion operators%
\begin{eqnarray}
\widetilde{a}_{a,j}^{\dag } &=&a_{A,j}^{\dag },  \notag \\
\widetilde{a}_{a,M+l}^{\dag } &=&\cos \theta e^{-i\varphi
_{B}^{l}}a_{B,l}^{\dag }+\sin \theta e^{i\varphi _{C}^{l}}a_{C,l}^{\dag },
\notag \\
\widetilde{a}_{a,M+N+l}^{\dag } &=&i(-1)^{n}[\sin \theta e^{-i\varphi
_{B}^{l}}a_{B,l}^{\dag }-\cos \theta e^{i\varphi _{C}^{l}}a_{C,l}^{\dag }],
\notag \\
&&
\end{eqnarray}%
where $\ j\in \lbrack 1,M]$, $l\in \lbrack 1,N]$. Similarly, when we set%
\begin{equation}
\phi =\frac{n}{2}+\frac{1}{4};t_{nB}=t\cos \theta ,t_{nC}=t\sin \theta
\end{equation}%
the Hamiltonian becomes%
\begin{equation}
H=\widetilde{H}_{a},
\end{equation}%
with $N_{a}=M+2N,$\ which is illustrated schematically in Fig. \ref{Q}(c).
Then we conclude that, when the flux is $\phi =n/2+1/4$ and the joint
hopping integrals satisfy $\sqrt{t_{nC}^{2}+t_{nB}^{2}}=t$, the QTBN is
equivalent to a single virtual open chain $a$ with length $M+2N$.

\subsubsection{Case: arbitrary $\protect\phi ,$ $t_{nB}=t_{nC}=\frac{t}{%
\protect\sqrt{2}}$}

When we take the interchain hopping integrals as $t_{nB}=t_{nC}=t/\sqrt{2}$,
the mapping (\ref{tildeQ}) reduces the network Hamiltonian as

\begin{eqnarray}
H &=&\widetilde{H}_{a}+\widetilde{H}_{b}+\widetilde{H}_{ab},  \notag
\label{film} \\
\widetilde{H}_{ab} &=&-\mu (\widetilde{a}_{a,M+N}^{\dag }\widetilde{a}%
_{a,M+N}-\widetilde{a}_{b,N}^{\dag }\widetilde{a}_{b,N})  \notag \\
&&-\widetilde{t}(\widetilde{a}_{b,N}^{\dag
}\widetilde{a}_{a,M+N}+H.c.).
\end{eqnarray}%
with $N_{a}=M+N$, $N_{b}=N$. We have replaced $i\widetilde{a}_{b,N}^{\dag }$
by $\widetilde{a}_{b,N}^{\dag }$ for $j\in \lbrack 1,N]$ without influence
on the physics of dynamical process.\ Here, $\widetilde{H}_{a}$ and $%
\widetilde{H}_{b}$ stand for two virtual chains with length $M+N$ and $N$
respectively, $\widetilde{H}_{ab}$\ represents the chemical potential at the
ends of chains $a$ and $b$ and the connection between the two sites. Note
that the end-site chemical potentials possess the same magnitude $\mu =t\cos
\Phi $, but of opposite sign and the hopping integral between the two end
sites is $\widetilde{t}=t\sin \Phi $. The reduced model is also illustrated
in Fig. \ref{Q}(d).

Physically, the chemical potentials $\mu $ and the hopping integral $%
\widetilde{t}\ $have the complementary relation $\mu ^{2}+\widetilde{t}%
^{2}=t^{2}$. When $\Phi =(n+1/2)\pi $, the network is equivalent to a linear
chain with length $M+2N$;\ while for $\Phi =n\pi $ it corresponds to two
independent chains with lengths $M+N$ and $N$. In the next section, we will
focus on such system for arbitrary $\Phi $. We will show that such system
behaves as an optical device, a transmission-reflection film, while the flux
determines the coefficient. In conclusion, the magnetic flux $\phi $ can
influence the \textquotedblleft effective length\textquotedblright\ or the
connective status of the virtual chains and then can be used to control the
dynamics of the network.

\subsection{Transmission-reflection film}

\begin{figure}[tbp]
\includegraphics[bb=30 195 560 700, width=7 cm,clip]{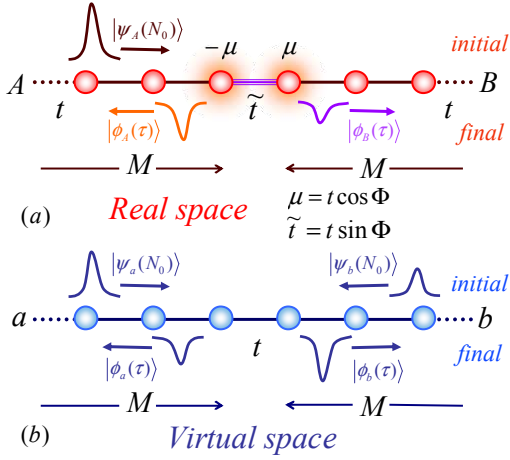}
\caption{\textit{(Color on line) (a) Schematic illustration for the
transmission-reflection film in real space. Two terminal sites of the two
TBNs $\{A,B\}$ are connected by the hopping integral $\widetilde{t}=t\sin
\Phi $. There also exist chemical potentials $\protect\mu =-t\cos \Phi $ for
the terminal site of chain $A$ and $\protect\mu =t\cos \Phi $ for the one of
chain $B$. (b) The above TBN can be recomposed as a homogeneous chain $%
\{a,b\}$ in virtual space.} }
\label{half}
\end{figure}

To make the above observation more clear, we consider two identical
tight-binding chains $\{A,B\}$, which consist of $N$ sites respectively.
There exists a connection interaction between the two terminal sites of the
two virtual chains. The hopping constant is $t\sin \Phi $ and each terminal
site also has a chemical potential, $\pm t\cos \Phi $. The Hamiltonian reads

\begin{equation}
H=H_{A}+H_{B}+H_{joint}  \label{Htr}
\end{equation}%
where $N_{A}=N_{B}=N$, $t_{j}^{[A]}=t_{j}^{[B]}=t$, and%
\begin{eqnarray}
H_{joint} &=&-t\sin \Phi (a_{A,N}^{\dag }a_{B,N}+H.c.)  \notag \\
&&-t\cos \Phi a_{A,N}^{\dag }a_{A,N}+t\cos \Phi a_{B,N}^{\dag }a_{B,N}.
\notag \\
&&
\end{eqnarray}%
Obviously, it is the simplest case of the system described by Eq. (\ref{film}%
).

In order to study the properties of this Bloch electron model more clearly,
we introduce two anticommutative sets of fermion operators%
\begin{eqnarray}
\widetilde{a}_{a,j}^{\dag } &=&\frac{\sqrt{2}}{2}\left( f_{+}a_{A,j}^{\dag
}-f_{-}a_{B,j}^{\dag }\right)  \notag \\
\widetilde{a}_{b,j}^{\dag } &=&\frac{\sqrt{2}}{2}\left( f_{-}a_{A,j}^{\dag
}+f_{+}a_{B,j}^{\dag }\right)  \label{tran}
\end{eqnarray}%
where $j\in \lbrack 1,N]$ and%
\begin{equation}
f_{\pm }=\cos \frac{\Phi }{2}\pm \sin \frac{\Phi }{2}.
\end{equation}%
The inverse transformation of Eq. (\ref{tran}) results in the reduction of
the network described by%
\begin{equation}
H=\widetilde{H}_{a}+\widetilde{H}_{b}+\widetilde{H}_{joint}  \label{ab}
\end{equation}%
where $N_{a}=N_{b}=N$, and%
\begin{equation}
\widetilde{H}_{joint}=-t(\widetilde{a}_{a,N}^{\dag }\widetilde{a}%
_{b,N}+H.c.).
\end{equation}%
Obviously, the Hamiltonian (\ref{ab}) depicts an imaginary linear chain with
homogeneous couplings $t$ no matter how much the magnitude of the flux $\Phi
$\ is taken. Then, the Hamiltonian describing transmission-reflection is
mapped into a chain in virtual space. Interestingly, such mapping is
irrelevant to the state concerned.

In order to demonstrate the function of such network, we study the
propagation of a moving GWP. To this end, we consider a GWP defined as (\ref%
{GWP}) at chain $A$, i.e.,%
\begin{equation}
\left\vert \psi _{A}(N_{0})\right\rangle =\frac{1}{\sqrt{\Omega _{1}}}%
\sum_{j=1}^{N}e^{-\frac{\alpha ^{2}}{2}(j-N_{0})^{2}}e^{i\frac{\pi }{2}%
j}a_{A,j}^{\dag }\left\vert 0\right\rangle .
\end{equation}%
We require this wave packet to satisfy $\left\langle 0\right\vert
a_{B,j}\left\vert \psi _{A}(N_{0})\right\rangle \simeq 0$,\ so that it
ensures the initial GWP being located in chain $A$. The transformation (\ref%
{tran}) means that such GWP corresponds to the combination of two GWPs in
virtual space with the centers at $N_{0}$ respectively,%
\begin{equation}
\left\vert \psi _{a(b)}(N_{0})\right\rangle =\frac{f_{+(-)}}{\sqrt{2\Omega
_{1}}}\sum_{j=1}^{N}e^{-\frac{\alpha ^{2}}{2}(j-N_{0})^{2}}e^{i\frac{\pi }{2}%
j}a_{a(b),j}^{\dag }\left\vert 0\right\rangle .
\end{equation}

Based on the analytical result in Ref. \cite{YS1}, the two GWPs in the
virtual chain should travel along the chain defined by Eq. (\ref{ab})\
without spreading as time evolution. Then, at a certain time $\tau $, the
evolved state $\left\vert \phi _{a(b)}(\tau )\right\rangle =$ $\exp (-i%
\widetilde{H}\tau )\left\vert \psi _{a(b)}(N_{0})\right\rangle $, or

\begin{eqnarray}
\left\vert \phi _{a(b)}(\tau )\right\rangle &=&\frac{1}{\sqrt{2\Omega _{1}}}%
[f_{+(-)}\sum_{j=1}^{N}e^{-\frac{\alpha ^{2}}{2}(j-N_{\tau })^{2}}e^{i\frac{%
\pi }{2}j}  \notag \\
&&+f_{-(+)}\sum_{j=1}^{N}e^{-\frac{\alpha ^{2}}{2}(Pj-N_{\tau })^{2}}e^{i%
\frac{\pi }{2}Pj}]a_{a(b),j}^{\dag }\left\vert 0\right\rangle  \notag \\
&&
\end{eqnarray}%
describes the superposition of two GWPs. Here, $Pj=2N+1-j,$ $N_{\tau
}=N_{0}+2t\tau $.

Transforming back to the real space, we rewrite the time evolution by the
state

\begin{equation}
\left\vert \Psi (\tau )\right\rangle =\cos \Phi \left\vert \phi _{A}(\tau
)\right\rangle +\sin \left\vert \phi _{B}(\tau )\right\rangle
\end{equation}%
in terms of the two components of the wave function%
\begin{eqnarray}
\left\vert \phi _{A}(\tau )\right\rangle &=&\frac{1}{\sqrt{\Omega _{1}}}%
\sum_{j=1}^{N}[e^{-\frac{\alpha ^{2}}{2}(Pj-N_{\tau })^{2}}e^{i\frac{\pi }{2}%
Pj}  \notag \\
&&+\frac{1}{\cos \Phi }e^{-\frac{\alpha ^{2}}{2}(j-N_{\tau })^{2}}e^{i\frac{%
\pi }{2}j}]a_{A,j}^{\dag }\left\vert 0\right\rangle  \notag \\
&\approx &\frac{1}{\sqrt{\Omega _{1}}}\sum_{j=1}^{N}e^{-\frac{\alpha ^{2}}{2}%
(Pj-N_{\tau })^{2}}e^{i\frac{\pi }{2}Pj}a_{A,j}^{\dag }\left\vert
0\right\rangle  \notag \\
&&  \label{phia}
\end{eqnarray}%
and

\begin{equation}
\left\vert \phi _{B}(\tau )\right\rangle =\frac{1}{\sqrt{\Omega _{1}}}%
\sum_{j=1}^{N}e^{-\frac{\alpha ^{2}}{2}(Pj-N_{\tau })^{2}}e^{i\frac{\pi }{2}%
Pj}a_{B,j}^{\dag }\left\vert 0\right\rangle .
\end{equation}%
Here, in Eq. (\ref{phia}), the second term is ignored in the case $%
\left\vert j-N_{\tau }\right\vert \gg 1$.

Obviously, the central positions of the final sub-GWPs $2N+1-N_{\tau }$
decrease with time $\tau $. This observation indicates that, the beam
splitter can split the GWP into two cloned GWPs with opposite moving
directions along with $AB$ chain. Therefore, state $\left\vert \phi
_{A}(\tau )\right\rangle $ represents the reflection component with
probability $\cos ^{2}\Phi $, while state $\left\vert \phi _{B}(\tau
)\right\rangle $\ is the transmission component through the connection of $%
AB $\ with probability $\sin ^{2}\Phi $. So this Bloch electron network for
a moving GWP behaves like an optical transmission-reflection film for
photons. Interestingly, transmission and reflection coefficients are
governed by the parameter $\Phi $, the flux through the network. This
feature is illustrated in Fig. \ref{half} in details.

\subsection{The dynamic properties of the $Q$-shaped Bloch electron model}

Now we take the propagation of the GWP $\left\vert \psi
_{A}(N_{0})\right\rangle $ as an example. Its advantage is that the GWP we
often used can move along a homogeneous chain without spreading
approximately. So we can easily see the various characteristics of the model
through the propagation of the GWP.

\subsubsection{Case: $\protect\phi =\frac{n}{2}$, $t_{nB}=t_{nC}=\frac{t}{%
\protect\sqrt{2}}$}

The initial GWP is moving with speed $2t$ along chain $A$. Before it reaches
the node, it can also be regarded as moving along the virtual chain $a$.
From the above discussion, the virtual chain $a$ is homogeneous with length $%
M+N$ and decoupled with another virtual chain $b$. Thus in the virtual
space, we can see that\ the GWP moves toward the end site of virtual chain $%
a $ and then reflects at the boundaries with \textquotedblleft $\pi $-phase
shift\textquotedblright . It never appears on virtual chain $b$. Notice
that, in this case, $t_{nB}=t_{nC}=t/\sqrt{2}$ must be satisfied, and then
the GWP in virtual space can be mapped into two identical GWPs with half
amplitude of the initial one in the real space. Therefore, the whole
propagation process in the real space is as follows: When the initial GWP
reaches the node, it is divided into the two identical GWPs which also move
with speed $2t$ along the legs $B$ and $C$ respectively without spreading.
Then the two GWPs reflect completely at the opposite site of the node and
come back along the original paths. When they reach the node again, they
merge as a big GWP and get out of the ring.

\subsubsection{Case: $\protect\phi =\frac{n}{2}+\frac{1}{4}$, $%
t_{nB}^{2}+t_{nC}^{2}=t^{2}$.}

As it is shown in the subsection (2), the reduction of $Q$-shaped Bloch
electron model have two mainly characters. First and foremost, it is
decoupled as a long virtual chain $a$ with length $M+2N$. Secondly, $%
t_{nB}=t\cos \theta $, $t_{nC}=t\sin \theta $. So when the initial GWP
reaches the node for the first time, it is divided into two GWPs with $%
\left( t_{nB}/t\right) ^{2}$ and $\left( t_{nC}/t\right) ^{2}$ amplitude of
the initial one. They move along the ring for one circle and reach the node
again. This time, instead of going out of the ring to the real chain $A$,
they reflect back and continue moving along the ring for another circle
until they meet at the node for the third time. After circumambulating two
circles, they finally merge into a big one and run out of the ring towards
to the input leg.

\subsubsection{Case: arbitrary $\protect\phi $, $t_{nB}=t_{nC}=\frac{t}{%
\protect\sqrt{2}}$}

For other values of $\phi $ and $t_{nB},t_{nC}$, when the two GWPs reach the
node for the second time, parts of them get out while the rest parts remain
moving in the ring. Especially, when $t_{nB}=t_{nC}=t/\sqrt{2}$, the
coupling constants and the chemical potentials satisfy the relation
discussed in the section \textquotedblleft Transmission-reflection
film\textquotedblright . So when the initial GWP reaches the end of the
virtual chain $a$, some novel phenomena occur. Part of it can move onto the
virtual chain $b$ and forms a new GWP with $\sin ^{2}\Phi $ amplitude of the
initial one. At the same time, the other part is reflected by the joint and
forms a GWP with $\cos ^{2}\Phi $ amplitude of the initial one. On mapping
them to the real space, we can image that when the two sub-GWPs reach the
node again. Parts of them are merged as a GWP with $\cos ^{2}\Phi $
amplitude getting out of the ring. However, the rest parts move along the
ring for another circle before going out.

Therefore, the magnetic flux $\phi $ can control the amplitude of the out
coming GWP. Such an $Q$-shaped Bloch electron model can also be used to test
the flux $\phi $ by measuring the probability of the out coming GWP at some
certain instants.

\section{Flux-controlled Interferometer and its reduction}

\subsection{Model and Hamiltonian}

\begin{figure}[tbp]
\includegraphics[bb=55 85 550 780, width=7 cm,clip]{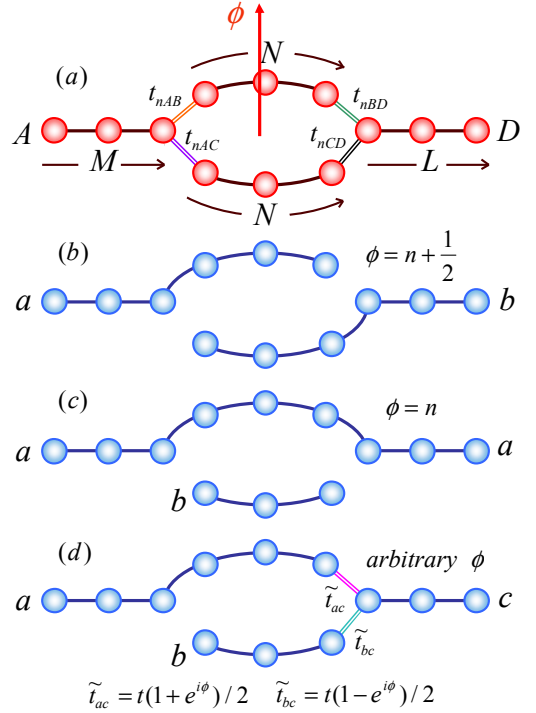}
\caption{\textit{(Color on line) (a) The $\protect\phi $-shaped Bloch
electron network with an input chain $A$, an output chain $D$ and a ring ${%
B,C}$ threaded by a magnetic flux. (b) When $t_{nAB}=t_{nCD}=t\cos \protect%
\theta $, $t_{nAC}=t_{nBD}=t\sin \protect\theta $ and $\protect\phi =n+1/2$,
the $\protect\phi $-shaped Bloch electron network can be decoupled into two
virtual homogeneous linear chains $a$ and $b$ with length $M+N$ and $N+L$
respectively. (c) When $t_{nAB}=t_{nBD}=t\cos \protect\theta $, $%
t_{nAC}=t_{nCD}=t\sin \protect\theta $ and $\protect\phi =n$, the $\protect%
\phi $-shaped Bloch electron network can be decoupled into a long virtual
homogeneous linear chain $a$ with length $M+N+L$ and a short virtual
homogeneous linear chain $b$ with length $N$. (d) If $%
t_{nAB}=t_{nAC}=t_{nBD}=t_{nCD}=t/\protect\sqrt{2}$, for an arbitrary $%
\protect\phi $, the network can be decoupled into three virtual homogeneous
linear chains $a$, $b$ and $c$. They connect with each other by the hopping
integrals $\widetilde{t}_{ac}$ and $\widetilde{t}_{bc}$. }}
\label{phai}
\end{figure}

In this subsection, we consider the interferometer model $\{A,B,C,D\}$ for
Bloch electron, illustrated schematically in Fig. \ref{phai}(a). This
quantum interferometer consists of two chains $A,D$ and one ring $\{B,C\}$
with one end of each chain connecting to two opposite point of the ring. The
ring is threaded by a magnetic flux $\phi $ in the unit of flux quanta. The
Hamiltonian reads
\begin{equation}
H=H_{A}+H_{D}+H_{B}+H_{C}+H_{joint}  \label{Hphi}
\end{equation}%
where $N_{A}=M$, $N_{D}=L$, $N_{B}=$ $N_{C}=N$, $t_{j}^{[A]}=$ $%
t_{j}^{[D]}=t $, $t_{j}^{[B]}=$ $t\exp (i\Phi _{B,j+1})$, and $t_{j}^{[C]}=$
$t\exp (-i\Phi _{C,j+1})$. The connection Hamiltonian reads%
\begin{eqnarray}
H_{joint} &=&-(t_{nAB}a_{A,M}^{\dag }a_{B,1}e^{i\Phi
_{B,1}}+t_{nAC}a_{A,M}^{\dag }a_{C,1}e^{-i\Phi _{C,1}}  \notag \\
&&+t_{nBD}a_{B,N}^{\dag }a_{D,1}e^{i\Phi _{B,N+1}}+  \notag \\
&&t_{nCD}a_{C,N}^{\dag }a_{D,1}e^{-i\Phi _{C,N+1}}+H.c.).
\end{eqnarray}%
Here, $\Phi =$ $\sum_{l=1}^{N+1}\Phi _{B,l}$ $+\sum_{l=1}^{N+1}\Phi _{C,l}$ $%
=2\pi \phi $ is the sum of $\Phi $ along the ring.

In this section, we investigate the basic properties of the flux-controlled
interferometer. Similarly, we will find out that there still exist some
analytical results, which reveal the dynamics of such network for the
appropriate parameters.

\subsection{Reduction of the interferometer network}

To reduce the network of interferometers, the four sets of new fermion
operator%
\begin{eqnarray}
\widetilde{a}_{a,j}^{\dag } &=&a_{A,j}^{\dag },  \notag \\
\widetilde{a}_{a,M+l}^{\dag } &=&\cos \theta e^{-i\varphi
_{B}^{l}}a_{B,l}^{\dag }+\sin \theta e^{i\varphi _{C}^{l}}a_{C,l}^{\dag },
\notag \\
\widetilde{a}_{b,l}^{\dag } &=&\sin \theta e^{-i\varphi
_{B}^{l}}a_{B,l}^{\dag }-\cos \theta e^{i\varphi _{C}^{l}}a_{C,l}^{\dag },
\notag \\
\widetilde{a}_{c,s}^{\dag } &=&e^{i\varphi _{C}^{N+1}}a_{D,s}^{\dag },
\label{Tran2}
\end{eqnarray}%
for $j\in \lbrack 1,M]$,\ $l\in \lbrack 1,N]$,\ and $s\in \lbrack 1,L],$\
are introduced to satisfy%
\begin{equation}
\left\{ \widetilde{a}_{a,M+j},\widetilde{a}_{b,j}^{\dag }\right\} =0.
\end{equation}%
Here,
\begin{equation}
\varphi _{\alpha }^{j}=\sum_{l=1}^{j}\Phi _{\alpha ,l},(\alpha =B,C),
\label{sumphase}
\end{equation}%
is the sum of the phase.

The inverse transformation of the above Eqs. (\ref{Tran2}) reduces the
Hamiltonian (\ref{Hphi}) into%
\begin{eqnarray}
H &=&\widetilde{H}_{a}+\widetilde{H}_{b}+\widetilde{H}_{d}  \notag \\
&&-t\sum_{j=1}^{N-1}(\widetilde{a}_{a,M+j}^{\dag }\widetilde{a}%
_{a,M+j+1}+H.c.)  \notag \\
&&-[(t_{nAB}\cos \theta +t_{nAC}\sin \theta )\widetilde{a}_{a,M}^{\dag }%
\widetilde{a}_{a,M+1}  \notag \\
&&+(-t_{nAB}\sin \theta +t_{nAC}\cos \theta )\widetilde{a}_{a,M}^{\dag }%
\widetilde{a}_{b,1}  \notag \\
&&+(t_{nBD}\cos \theta e^{i\Phi }+t_{nCD}\sin \theta )\widetilde{a}%
_{a,M+N}^{\dag }\widetilde{a}_{c,1}  \notag \\
&&+(-t_{nBD}\sin \theta e^{i\Phi }+t_{nCD}\cos \theta )\widetilde{a}%
_{b,N}^{\dag }\widetilde{a}_{c,1}+H.c.]  \notag \\
&&
\end{eqnarray}%
where $N_{a}=M$, $N_{b}=N$, and $N_{c}=L$. Now we concentrate on two special
cases with different $\phi $ and other parameters:

\subsubsection{Case: $\protect\phi =n+\frac{1}{2}$, $t_{nAB}=t_{nCD}=t\cos
\protect\theta $, $t_{nAC}=t_{nBD}=t\sin \protect\theta $}

It is obvious that $\exp (i\Phi )=-1$. The Hamiltonian can be rewritten as%
\begin{eqnarray}
H &=&-t(\sum_{j=1}^{M+N-1}\widetilde{a}_{a,j}^{\dag }\widetilde{a}%
_{a,j+1}+\sum_{j=1}^{L-1}\widetilde{a}_{c,j}^{\dag }\widetilde{a}_{c,j+1}
\notag \\
&&+\sum_{j=1}^{N-1}\widetilde{a}_{b,j}^{\dag }\widetilde{a}_{b,j+1}+%
\widetilde{a}_{b,N}^{\dag }\widetilde{a}_{c,1}+H.c.).  \notag \\
&&
\end{eqnarray}%
We define the new fermion operator%
\begin{equation}
\widetilde{a}_{b,N+j}^{\dag }=\widetilde{a}_{c,j}^{\dag },(j\in \lbrack 1,L])
\end{equation}%
to extend the virtual chain $b$. Its Hamiltonian
\begin{equation}
H=\widetilde{H}_{a}+\widetilde{H}_{b}  \label{Hphir}
\end{equation}%
is given by the parameters $N_{a}=M+N$ and $N_{b}=N+L$.

From the reduced Hamiltonian (\ref{Hphir}), we can see that the
interferometer network is decoupled into two imaginary linear chains with
homogeneous couplings $t$. The set of $\{\widetilde{a}_{a,j}^{\dag }\mid
j\in \lbrack 1,N+M]\}$ defines one of them with length $M+N$ sites, and the
set $\{\widetilde{a}_{a,j}^{\dag }\mid j\in \lbrack 1,N+L]\}$\ defines the
other one with length $N+L$. As illustrated schematically in Fig. \ref{phai}%
(b).

\subsubsection{Case: $\protect\phi =n$, $t_{nAB}=t_{nBD}=t\cos \protect%
\theta $, $t_{nAC}=t_{nCD}=t\sin \protect\theta $}

In this case, the Hamiltonian becomes%
\begin{eqnarray}
H &=&-t(\sum_{j=1}^{M+N-1}\widetilde{a}_{a,j}^{\dag }\widetilde{a}%
_{a,j+1}+\sum_{j=1}^{L-1}\widetilde{a}_{c,j}^{\dag }\widetilde{a}_{c,j+1}
\notag \\
&&+\sum_{j=1}^{N-1}\widetilde{a}_{b,j}^{\dag }\widetilde{a}_{b,j+1}+%
\widetilde{a}_{a,M+N}^{\dag }\widetilde{a}_{c,1}+H.c.)  \notag \\
&&
\end{eqnarray}%
with the newly defined operators%
\begin{equation}
\widetilde{a}_{a,M+N+j}^{\dag }=\widetilde{a}_{c,j}^{\dag },(j\in \lbrack
1,L])
\end{equation}%
the reduced Hamiltonian

\begin{equation}
H=\widetilde{H}_{a}+\widetilde{H}_{b}
\end{equation}%
describe an extended virtual chain of length $N_{a}=M+N+L$ and another of $%
N_{b}=N$.

It is clear that, when the conditions
\begin{eqnarray}
t_{nAB} &=&t_{nBD}=t\cos \theta  \notag \\
t_{nAC} &=&t_{nCD}=t\sin \theta  \notag \\
\phi &=&n
\end{eqnarray}%
are satisfied, the interferometer network is decoupled into two imaginary
linear chains with length $M+N+L$ sites and $N$ sites respectively. See also
Fig. \ref{phai}(c).

\subsubsection{Case: arbitrary $\protect\phi $, $t_{nAB}=t_{nAC}=$ $%
t_{nBD}=t_{nCD}$ $=\frac{t}{\protect\sqrt{2}}$}

Under this condition, the Hamiltonian is reduced as%
\begin{equation}
H=\widetilde{H}_{a}+\widetilde{H}_{b}+\widetilde{H}_{c}+\widetilde{H}_{joint}
\end{equation}%
where $N_{a}=M+N$, $N_{b}=N$, and $N_{c}=L$.

Here, the joint Hamiltonian is%
\begin{eqnarray}
\widetilde{H}_{joint} &=&-t[e^{i\frac{\Phi }{2}}\cos \left( \frac{\Phi }{2}%
\right) \widetilde{a}_{a,M+N}^{\dag }\widetilde{a}_{c,1}  \notag \\
&&-ie^{i\frac{\Phi }{2}}\sin \left( \frac{\Phi }{2}\right) \widetilde{a}%
_{b,N}^{\dag }\widetilde{a}_{c,1}+H.c.].
\end{eqnarray}%
while the sub-Hamiltonians $\widetilde{H}_{a}$, $\widetilde{H}_{b}$ and $%
\widetilde{H}_{c}$ present three homogeneous virtual linear chains $\{a,b,c\}
$ with length $M+N$, $N$ and $L$ respectively. In $\widetilde{H}_{joint}$,
there exists a connection interaction $\exp (i\Phi /2)\cos \left( \Phi
/2\right) $ between the two end sites of virtual chain $a$ and $c$.
Meanwhile, there exists another connection interaction $-i\exp (i\Phi
/2)\sin (\Phi /2)$ between the two end sites of virtual chain $b$ and $c$.
The geometry of such network is illustrated in Fig. \ref{phai}(d). Obviously
in the virtual space, such network is equivalent to the $Y$-shaped beam
splitter with different lengths of output arms and complex joint hopping
constants controlled by the flux $\Phi $. From the discussion about $Y$%
-shaped network, we have known that the lengths of output arms do not affect
the feature as beam splitter for local input wave packet. In the following
we will investigate property of such $Y$-shaped network by considering
equi-length case for simplicity.

\subsection{Y-shaped Beam splitter controlled by $\Phi $}

Now we consider a $Y$-shaped network $\{A,B,C\}$ with complex joint hopping
constants. The model Hamiltonian reads

\begin{equation}
H=H_{A}+H_{B}+H_{C}+H_{joint}
\end{equation}%
where $N_{A}=L$, $N_{B}=N_{C}=N$, and $t_{j}^{[A]}=$ $t_{j}^{[B]}=$ $%
t_{j}^{[C]}=t$. The joint Hamiltonian%
\begin{equation}
H_{joint}=-(t_{AB}a_{A,L}^{\dag }a_{B,1}+t_{AC}a_{A,L}^{\dag
}a_{C,1}+H.c.), \notag
\end{equation}%
describes the connections with the complex hopping integrals

\begin{equation}
t_{AB}=te^{-i\frac{\Phi }{2}}\cos \left( \frac{\Phi }{2}\right)
,t_{AC}=ite^{-i\frac{\Phi }{2}}\sin \left( \frac{\Phi }{2}\right) .
\end{equation}

Interestingly, if we get rid of the exponential terms in the hopping
integrals, i.e., $t\cos \left( \Phi /2\right) $, $t\sin \left( \Phi
/2\right) $, we recover the matching condition (\ref{matching}) in the
original $Y$-shaped beam splitter $t_{AB}^{2}+t_{AC}^{2}=t^{2}$. In order to
decouple this network, we introduce three communitative sets of fermion
operators%
\begin{eqnarray}
\widetilde{a}_{a,l}^{\dag } &=&a_{A,l}^{\dag },  \notag \\
\widetilde{a}_{a,L+j}^{\dag } &=&e^{i\frac{\Phi }{2}}[\cos \left( \frac{\Phi
}{2}\right) a_{B,j}^{\dag }-i\sin \left( \frac{\Phi }{2}\right)
a_{C,j}^{\dag }],  \notag \\
\widetilde{a}_{b,j}^{\dag } &=&e^{-i\frac{\Phi }{2}}[i\sin \left( \frac{\Phi
}{2}\right) a_{B,j}^{\dag }+\cos \left( \frac{\Phi }{2}\right) a_{C,j}^{\dag
}],  \notag \\
&&  \label{tran3}
\end{eqnarray}%
for $l\in \lbrack 1,L]$\ and $j\in \lbrack 1,N]$.

The inverse transformations of Eqs. (\ref{tran3}) result in the reduction of
Hamiltonian in terms of $\widetilde{a}_{a,j}^{\dag }$, $\widetilde{a}%
_{a,L+j}^{\dag }$, and $\widetilde{a}_{b,j}^{\dag }$:%
\begin{equation}
H=\widetilde{H}_{a}+\widetilde{H}_{b}
\end{equation}%
where $N_{a}=L+N$ and $N_{b}=N$.

Thus this kind of $Y$-shaped Bloch electron network is also decoupled into
two imaginary chains. Similarly, we apply the beam splitter to the special
Bloch electron GWP $\left\vert \psi _{A}(N_{0})\right\rangle $. At a certain
time $\tau $, such GWP evolves into%
\begin{eqnarray}
\left\vert \Psi (\tau )\right\rangle &\sim &\cos \left( \frac{\Phi }{2}%
\right) \left\vert \psi _{B}(N_{\tau })\right\rangle  \notag \\
&&-i\sin \left( \frac{\Phi }{2}\right) \left\vert \psi _{C}(N_{\tau
})\right\rangle ,
\end{eqnarray}%
where $N_{\tau }=N_{0}+2t\tau -L$. This means that the beam splitter can
split the GWP into two cloned GWPs completely with the probabilities%
\begin{eqnarray}
\left\vert \left\langle j_{B}\right. \left\vert \psi _{B}(N_{\tau
})\right\rangle \right\vert ^{2} &=&\cos ^{2}\frac{\Phi }{2};  \notag \\
\left\vert \left\langle j_{C}\right. \left\vert \psi _{C}(N_{\tau
})\right\rangle \right\vert ^{2} &=&\sin ^{2}\frac{\Phi }{2},
\end{eqnarray}%
which can be controlled by the external flux $\phi $.

\subsection{AB effect in a solid system}

This virtual model of interferometer network is very similar to the second
type of $Y$-shaped beam splitter we discussed before. The only difference
between them is that in this virtual model the lengths of the two legs are
unequal. Fortunately, by appropriately choosing $\alpha $, the width of the
wave packet, the GWP can be regarded as a classical electron. It not only
propagates along a homogeneous chain without spreading approximately, but
also does not regard the length of the chain. Now we prepare such a moving
GWP at the input leg. When it reaches to the node, it will be split into two
cloned GWPs with the amplitudes of $\cos ^{2}(\Phi /2)$ and $\sin ^{2}(\Phi
/2)$ respectively. According to our discussion above, the sub-GWP with the
amplitude of $\sin ^{2}(\Phi /2)$ will be reflected by the opposite node of
the ring, but the other sub-GWP with the amplitude of $\cos ^{2}(\Phi /2)$
will move onto the output leg directly. So some time later we will receive a
cloned GWP with the probability $\cos ^{2}(\Phi /2)$ at the output leg.

\begin{figure}[tbp]
\includegraphics[bb=25 430 570 670, width=7 cm,clip]{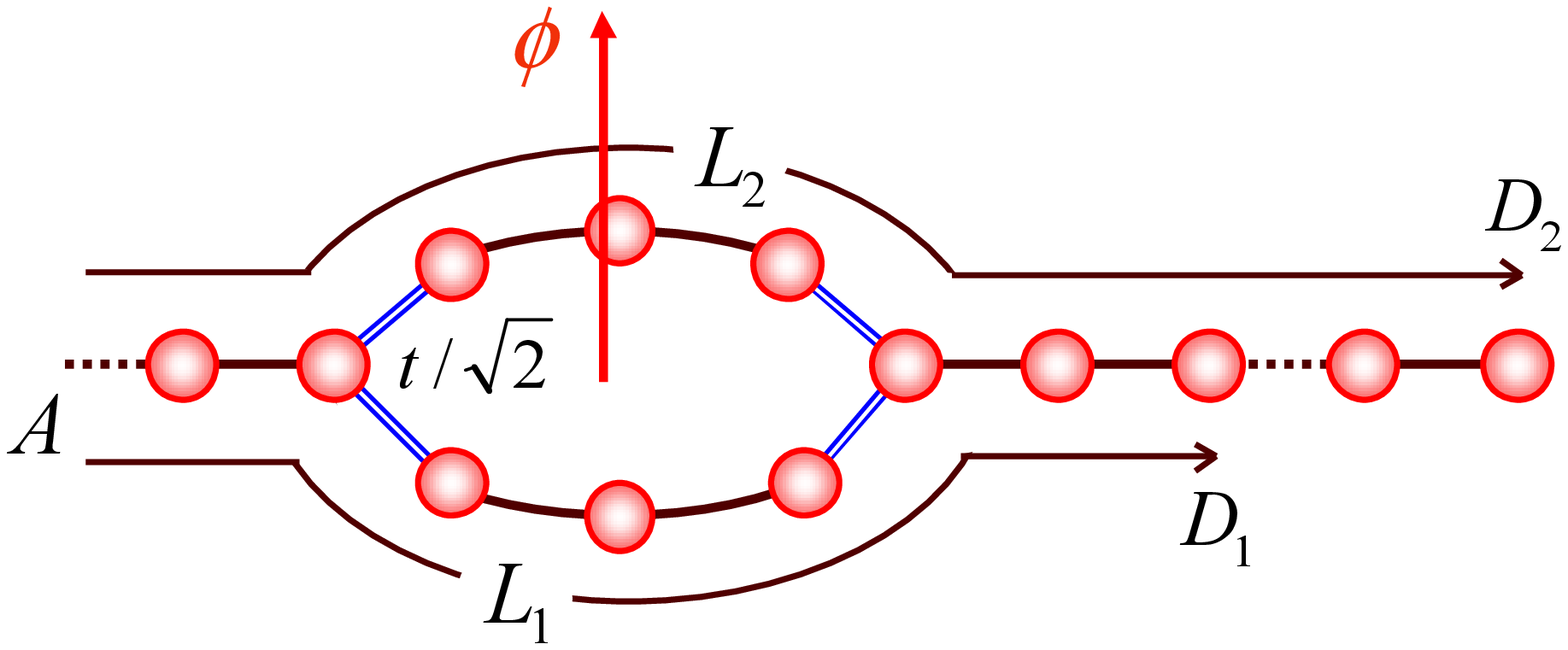} %
\includegraphics[bb=18 436 580 720, width=7 cm,clip]{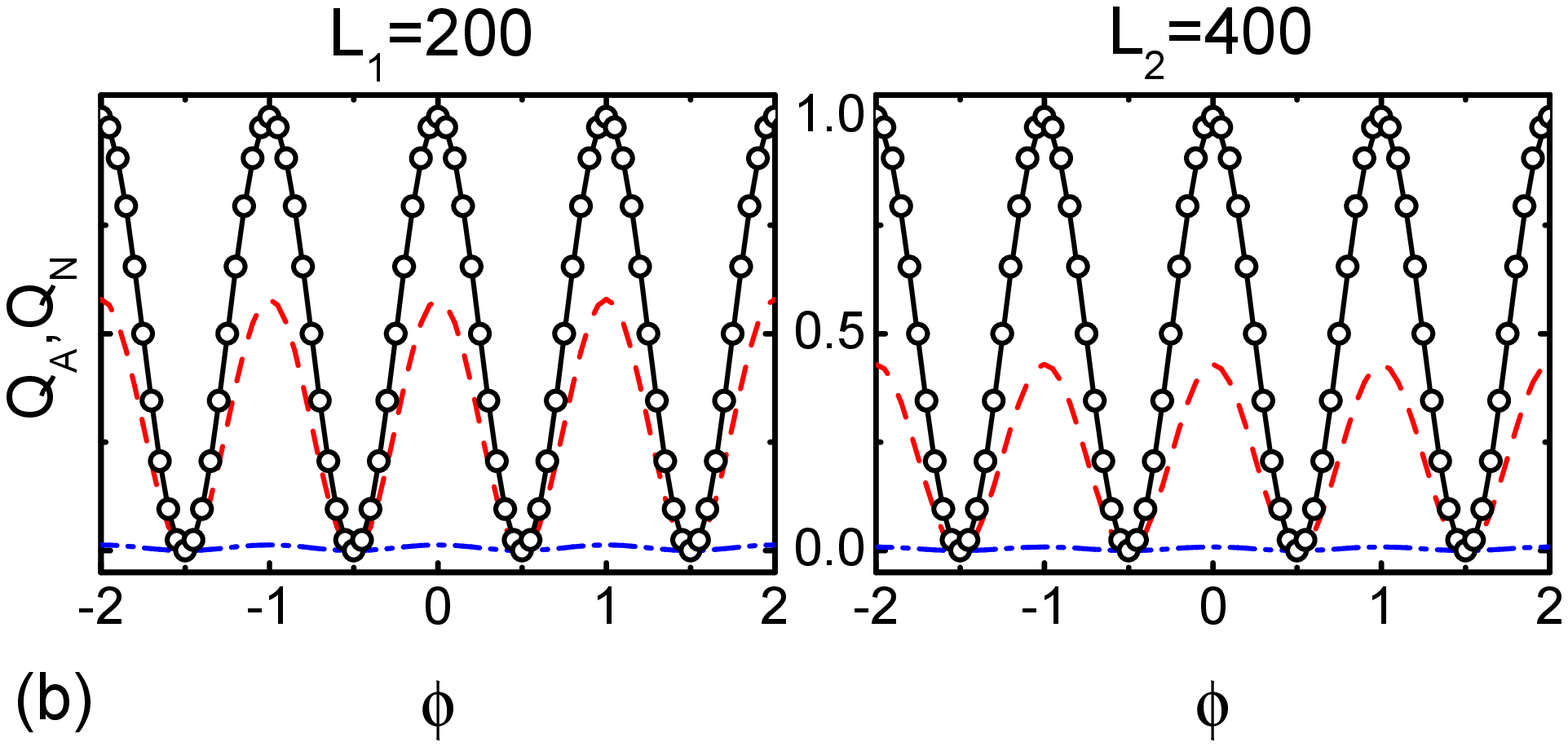}
\caption{\textit{(Color on line) (a) The schematic illustration for the
mimic A-B effect device in a solid system. (b) The comparison of relative
probability $Q$ as a function of the magnetic flux $\protect\phi $. The
half-width of the initial wave packets are 16.65 ($\protect\alpha =0.1$,
blue solid line), 5.55 ($\protect\alpha =0.3$, red dash line), 1 ($\protect%
\alpha =\infty $, black dot dash line). The optical paths between the input
site $A$ and the detector sites $D_{1}$, $D_{2}$ are $L_{1}=200$ (left), $%
L_{2}=400$ (right). It shows that the relative probability $Q$ are periodic
in the magnetic flux $\protect\phi $ with a period of unit flux quantum $%
\Phi _{0}=h/e$. When the half-width of the initial wave packet is narrower
or the optical path is longer, the relative probability $Q$ is smaller, the
A-B effect is weaker.}}
\label{AB}
\end{figure}

The interferometer based on Bloch electron network can be regarded as a
mimic of AB effect \cite{ABeffect} experimental device in a solid system
illustrated in Fig. \ref{AB}(a). Here, the initial GWP%
\begin{equation}
\left\vert \psi _{A}(N_{0})\right\rangle =\frac{1}{\sqrt{\Omega _{1}}}%
\sum_{j=1}^{M}e^{-\frac{\alpha ^{2}}{2}(j-N_{0})^{2}}e^{i\frac{\pi }{2}%
j}\left\vert j\right\rangle
\end{equation}%
is taken as a good example to demonstrate the physical mechanism of such
setup.

We first focus on the GWP at the input site $N_{0}=A$ and detected it later
on a distant site $D_{1}$ or $D_{2}$. The maximal probability of the GWP in
some certain site $j$ is
\begin{equation}
|\psi (j,\alpha )|_{\max }^{2}=\max \{\left\vert \left\langle j\right\vert
e^{-iH\tau }\left\vert \psi _{A}(N_{0})\right\rangle \right\vert ^{2}\}
\end{equation}%
Thus we can define the relative probability $Q$ as a function of $\alpha $,
the magnetic flux $\phi $, and the site of the detector $j$,%
\begin{equation}
Q(j,\phi ,\alpha )=\frac{|\psi (j,\alpha )|_{\max }^{2}}{|\psi (A,\alpha
)|_{\max }^{2}}.
\end{equation}

Obviously, $Q$ is an observable physical quantity, which describes the AB
effect and the influence of lattice scattering. Numerical simulation of $%
Q(D_{1},\phi ,\alpha )$ and $Q(D_{2},\phi ,\alpha )$ for different initial
GWP with half-width $\Delta =16.65$, $(\alpha =0.1)$, $\Delta =5.55$, $%
(\alpha =0.3)$, and $\Delta =1$, $(\alpha =\infty )$ are plotted in Fig. \ref%
{AB}(b). Here, the optical paths between the input site and the
detector-sites are $L_{1}=200$, $L_{2}=400$. The ring of the system is
threaded by a magnetic flux $\phi \in \lbrack -2,2]$. The numerical results
show that the relative probabilities $Q$\ are periodic in the magnetic flux $%
\phi $ with a period of unit flux quantum $\Phi _{0}=h/e$. This is the so
called AB effect in a solid system. At $\phi =$integer, our previous
discussion shows that the interferometer network of Bloch electron model is
decoupled into one long chain and one short chain. The initial GWP localized
in the input arm can be transmitted to the detector-arm without any
reflection. Consequently, the relative probability $Q$ reaches its maximum
of the curve. On the other hand, when $\phi $ is a half-integer, the initial
GWP cannot be transmitted to the detector-arm. Thus, the corresponding $Q$
equals to its minimum zero.

Then we consider the GWPs with different half-width $\Delta =2\sqrt{\ln 2}%
/\alpha $. If $\Delta $ is larger, the GWP is localized in the linear
dispersion regime more exactly. In this case, it can be well transferred
without spreading \cite{YS1}; or in another point of view, it is a free
particle which will not be scattered by the lattice. We can see from the
numerical results, $Q=$ $(1+\cos \Phi )/2$, the maximum $Q$ of $\Delta
=16.65 $, $(\alpha =0.1)$ is approximately equals to $1$. Otherwise, a GWP
with smaller $\Delta $, i.e., $\Delta =5.55$, $(\alpha =0.3)$ is scattered
by the lattice severely, so the relative probability $Q$ of which are
smaller than $Q$ of wider GWP. It is reasonable that when the optical path
is longer, the influence of lattice scattering is larger, the relative
probability $Q$ is smaller. The black dot dash line also shows that in large
$\alpha $ limit, $Q $ are approximately equal to zero. Therefore, a GWP
localized beyond the linear dispersion regime is not suitable for
demonstrating the AB effect experiment in a solid system.

To sum up, when the half-width of the initial wave packet is narrower or the
detect-length is longer, the relative probability $Q$ is smaller, the AB
effect is weaker to be observed. From these results, we get two possible
reasons why we cannot observe AB effect in a macroscopically solid system.
Firstly, we do not choose an appropriate wave packet. Secondly, the total
site of the macroscopically solid system is so large that the influence of
lattice scattering cannot be ignored. To solve these problems and to realize
AB effect in a solid system, we should choose a wider GWP mentioned in our
previous work \cite{YS1}. At the same time, we should decrease the optical
paths between the input site and the detector-sites.

\section{Applications for spin network}

In the above discussion, we studied the fermion systems where the Bloch
electrons move along the quantum lattice network. We consider various
geometrical configurations of TBN that are analogous to quantum optical
devices, such as beam splitters and interferometers. In this section, we
will apply the results obtained for TBN to another analogue system, spin
network (SN) where the spin wave acts as the Bloch electron.

The basic setup of a SN is constructed topologically by the linear spin
chains and the various connections between the ends of them. Here, we
consider the spin-$1/2$ $XY$ model, in which only the nearest neighbor (NN)
coupling term is taken into account. The Hamiltonian of a SN reads

\begin{equation}
H^{s}=\sum\limits_{l}H_{l}^{s}+H_{joint}^{s},  \label{Hs}
\end{equation}%
\ where the Hamiltonians of leg $l$ consisting of $N_{l}$ spins and the
joints are%
\begin{eqnarray}
H_{l}^{s} &=&H_{l}^{s}(J_{l},N_{l})\circeq
\sum_{j=1}^{N_{l}-1}J_{j}^{[l]}(S_{l,j}^{+}S_{l,j+1}^{-}+H.c.),  \notag \\
H_{joint}^{s} &\equiv &J_{ji}^{[lm]}S_{l,j}^{+}S_{m,i}^{-}+H.c..
\end{eqnarray}%
Here, $S_{l,j}^{\pm }$ are the Pauli spin operators acting on the internal
space of electron on the $j$th site of the $l$th leg. Although the SNs and
TBNs have the same structure, the physics should be different due to the
difference of the intrinsic statistical properties. Then the analytical
conclusions for TBN are not available to SN. However, in the context of
state transfer, only the dynamics of the single-magnon is concerned. Notice
that for Hamiltonian (\ref{Hs}), the $z$-component of the total spin $%
S^{z}=\sum\nolimits_{l,i}S_{l,i}^{z}$ is conserved, i.e. $[S^{z},H^{s}]=0$.
Thus in the invariant subspace with $S^{z}=(\sum\nolimits_{l}N_{l}-1)/2$,
this model can be mapped into single-particle TBN.

\section{Conclusion and remark}

In summary, we studied various geometrical configurations of tight-binding
networks\ for the fermion systems. It is found that the lattice networks for
moving GWPs are analogous to quantum optical devices, such as beam splitters
and interferometers. In practice, our coherent quantum network for
electronic wave can be implemented by an array of quantum dots, Josephson
junctions or other artificial atoms. It will enable an elementary quantum
device for scalable quantum computation, which can coherently transfer
quantum information among the integrated qubits.\textbf{\ }The observable
effects for electronic wave interference may be discovered in the dynamics
of magnetic domain in some artificial quantum material.

In the above studies, we only consider the spinless\ Bloch electron.
Actually, all the conclusions we obtained can be extended to the networks of
spin-$1/2$ electrons, if the external magnetic field does not exert any
forces or torques on the magnetic moment of spin, but only a phase on the
wave function of electron. The Hamiltonian of such system has the similar
form with (\ref{h}) and (\ref{joint}) under the transformation $a_{j}^{\dag
}a_{i}\longrightarrow \sum_{\sigma =\pm 1}a_{j,\sigma }^{\dag }a_{i,\sigma }$%
. Note that, for the new Hamiltonian, the spin of electron is a conservative
quantity that cannot be influenced during the propagation \cite{YS1}. The
electronic wave packet with spin polarization is an analogue of photon
\textquotedblleft flying qubit\textquotedblright , where the quantum
information was encoded in its two polarization states. Thus, these networks
can function as some optical devices, such as beam splitters and
interferometers. These are expected to be used as quantum information
processors for the fermion system based on the possible engineered solid
state systems, such as the array of quantum dots, Josephson junctions or
other artificial atoms that can be implemented experimentally.

This work is supported by the NSFC with grant Nos. 90203018, 10474104 and
60433050; and by the National Fundamental Research Program of China with
Nos. 2001CB309310 and 2005CB724508.


\begin{thebibliography}{99}
\bibitem{QIP1} D. Bouwnmeester, A. Ekert, A. Zeilinger(Eds.),
\textquotedblleft The Physics of Quantum Information\textquotedblright ,
Springer, Berlin, 2000.

\bibitem{QIP2} M.A. Nielsen , I.L. Chuang, \textquotedblright Quantum
Computation and Quantum Information\textquotedblright . Cambridge University
Press, Cambridge, U.K. 2000.

\bibitem{QST1} S. Bose, Phys. Rev. Lett. \textbf{91}, 207901 (2003); S.
Bose, B-Q. Jin and V. E. Korepin, Phys. Rev. A \textbf{72}, 022345 (2005);
S. Bose, Phys. Rev. Lett. \textbf{91}, 207901 (2003); M-H. Yung and S. Bose,
Phys. Rev. A \textbf{71}, 032310 (2005).

\bibitem{QST2} V. Subrahmanyam, Phys. Rev. A \textbf{69}, 034304 (2004).

\bibitem{QST3} M. Christandl, N. Datta, A. Ekert and A. J. Landahl, Phys.
Rev. Lett. \textbf{92}, 187902 (2004)

\bibitem{QST4} C. Albanese, M. Christandl, N. Datta and A. Ekert, Phys. Rev.
Lett. \textbf{93}, 230502 (2004)

\bibitem{QST5} T. J. Osborne and N. Linden, Phys. Rev. A \textbf{69}, 052315
(2004).

\bibitem{LY} Y. Li, T. Shi, B. Chen, Z. Song, C.P. Sun, Phys. Rev. A \textbf{%
71}, 022301 (2005).

\bibitem{ST} T. Shi, Y. Li, Z. Song, and C.P. Sun, Phys. Rev. A \textbf{71},
032309 (2005)

\bibitem{SZ} Z. Song, C.P. Sun, Low Temperature Physics \textbf{31}, 686
(2005).

\bibitem{QST6} M.B. Plenio, F. L. Semiao, New. J. Phys. \textbf{7}, 73 (2005)

\bibitem{quant-network} J. E. Avron; A. Raveh and B. Zur, Rev. Mod. Phys.
\textbf{60}, 873 (1988); C. H. Wu, G. Mahler, Phys. Rev. B \textbf{43}, 5012
(1991); J. Vidal, G. Montambaux and B. Doucot, Phys. Rev. B \textbf{62},
R16294 (2000); P. S. Deo and A. M. Jayannavar, Phys. Rev. B \textbf{50},
11629 (1994).

\bibitem{spin-network} A. Kay and M. Ericsson, New J. Phys. \textbf{7}, 143
(2005); G. D. Chiara, R. Fazio, C. Macchiavello, S. Montangero and G. M.
Palma, Phys. Rev. A \textbf{72}, 012328 (2005); M. Paternostro, G. M. Palma,
M. S. Kim and G. Falci, Phys. Rev. A \textbf{71}, 042311 (2005).

\bibitem{Experiment} R. A. Webb, S. Washburn, C. P. Umbach, and R. B.
Laibowitz, Phys. Rev. Lett. \textbf{54}, 2696 (1985); V. Chandrasekhar, M.
J. Rooks, S. Wind, and D. E. Prober, Phys. Rev. Lett. \textbf{55}, 1610
(1985).

\bibitem{Plenio} M.B. Plenio, J. Hartley and J. Eisert, New J. Phys. \textbf{%
6}, 36 (2004); A Perales, M.B. Plenio, J. Opt. B \textbf{7},
S601-S609 (2005).

\bibitem{gauge-trans} N. Byers and C.N. Yang, Phys. Rev. Lett. \textbf{7},
46 (1961);\ H. T. Nieh, G. Su, B-H. Zhao, Phys. Rev. B \textbf{51}, 3760
(1995).

\bibitem{tight-binding} D. Langbein, Phys. Rev. \textbf{180}, 633 (1969);
G-Y. Oh, J. Korean Phys. Soc. 42, 714 (2003).

\bibitem{Peierls} R. Peierls, Z. Physik \textbf{80}, 763 (1933).

\bibitem{YS1} S. Yang, Z. Song, and C.P. Sun, Phys. Rev. A \textbf{73},
022317 (2006).

\bibitem{QOP} R. Loudon, \textit{The quantum theory of light,} (Oxford,
2000); M.O. Scully and M.S. Zubairy, Quantum Optics, (Oxford, 1997).

\bibitem{S-photon} J.D. Franson; Phys. Rev. A \textbf{56}, 1800-1805 (1997);
K. Jacobs and P.L. Knight; Phys. Rev. A \textbf{54}, R3738(1996); T. Wang,
M. Kostrun, and S.F. Yelin; Phys. Rev. A \textbf{70}, 053822 (2004).

\bibitem{Entng} M. Zukowski, A. Zeilinger, and M.A. Horne, Phys. Rev. A
\textbf{55}, 2564(1997); J.L. van Velsen, Phys. Rev. A \textbf{72}, 012334
(2005).

\bibitem{N-interfer} H. Rauch, W. Treimer, and U. Bonse, Phys. Lett. A
\textbf{47}, 369 (1974).

\bibitem{C-atom} D. Cassettari, B. Hessmo, R. Folman, T. Maier, and J.
Schmiedmayer, Phys. Rev. Lett. \textbf{85}, 5483(2000); U. V. Poulsen and K.
Momer, Phys. Rev. A \textbf{65}, 033613 (2002); D. C. E. Bortolotti and J.L.
Bohn; Phys. Rev. A \textbf{69}, 033607 (2004).

\bibitem{BEC} F. Burgbacher and J. Audretsch; Phys. Rev. A \textbf{60},
R3385(1999); N.M. Bogoliubov, A. G. Izergin, N.A. Kitanine, A.G.
Pronko, and J. Timonen; Phys. Rev. Lett. \textbf{86}, 4439 (2001)

\bibitem{wang} X. Wang and P. Zanardi, Phys. Lett. A \textbf{301}, 1 (2002);
X. Wang, Phys. Rev. A \textbf{66}, 034302 (2002).

\bibitem{qian} X-F. Qian, Y. Li, Y. Li, Z. Song, and C.P. Sun, Phys. Rev. A
\textbf{72}, 062329 (2005).

\bibitem{ABeffect} Y. Aharonov and D. Bohm, Phys. Rev. \textbf{115}, 485
(1959).
\end{thebibliography}
\end{document}